\documentclass{elsarticle}
\pdfoutput=1
\usepackage{lineno}
\modulolinenumbers[5]
\usepackage{graphicx}
\graphicspath{{./}}
\usepackage{subcaption}
\usepackage{wrapfig}
\usepackage{array}
\usepackage{bm}
\usepackage{framed} 
\usepackage{multicol} 
\usepackage{tabularx}
    \newcolumntype{C}{>{\centering\arraybackslash}b{1.8cm}}
    \newcolumntype{D}{>{\centering\arraybackslash}b{2.1cm}}
\usepackage{booktabs}
\usepackage{multirow}
\usepackage[dvipsnames]{xcolor}

\usepackage[detect-none]{siunitx}
\usepackage{nomencl}
\renewcommand*\nompreamble{\begin{multicols}{2}}
\renewcommand*\nompostamble{\end{multicols}}
\makenomenclature

\usepackage{amsmath,amssymb}
\newsavebox\myboxA
\newsavebox\myboxB
\newlength\mylenA

\newcommand*\xoverline[2][0.75]{%
    \sbox{\myboxA}{$\math#2$}%
    \setbox\myboxB\null
    \ht\myboxB=\ht\myboxA%
    \dp\myboxB=\dp\myboxA%
    \wd\myboxB=#1\wd\myboxA
    \sbox\myboxB{$\math\overline{\copy\myboxB}$}
    \setlength\mylenA{\the\wd\myboxA}
    \addtolength\mylenA{-\the\wd\myboxB}%
    \ifdim\wd\myboxB<\wd\myboxA%
       \rlap{\hskip 0.5\mylenA\usebox\myboxB}{\usebox\myboxA}%
    \else
        \hskip -0.5\mylenA\rlap{\usebox\myboxA}{\hskip 0.5\mylenA\usebox\myboxB}%
    \fi}
    \makeatother
\usepackage[margin=2cm]{geometry}

\newcommand{\ubar}{\mathbf{\overline{u}}}

\usepackage{etoolbox}
\renewcommand\nomgroup[1]{%
  \item[\bfseries
  \ifstrequal{#1}{F}{Flow Parameters}{%
  \ifstrequal{#1}{D}{Foil Dimensions}{%
  \ifstrequal{#1}{K}{Kinematic Parameters}{%
  \ifstrequal{#1}{P}{Performance Metrics}{}}}}%
]}

\journal{Ocean Engineering}

\biboptions{authoryear}

\makeatletter
\def\ps@pprintTitle{%
 \let\@oddhead\@empty
 \let\@evenhead\@empty
 \def\@oddfoot{\footnotesize \copyright 2020. This manuscript version is made available under the CC-BY-NC-ND 4.0 license http://creativecommons.org/licenses/by-nc-nd/4.0/ \hfill}%
 \let\@evenfoot\@oddfoot}
\makeatother

\begin{document}

\begin{frontmatter}

\title{Variable thrust and high efficiency propulsion\\ with oscillating foils at high Reynolds numbers}




\author[uw-affiliation]{Mukul Dave\corref{mycorrespondingauthor}}
\cortext[mycorrespondingauthor]{Corresponding author} 
\ead{mhdave@wisc.edu}

\author[engg-affiliation]{Arianne Spaulding}
\ead{arianne\_spaulding@alumni.brown.edu}

\author[uw-affiliation]{Jennifer A. Franck}
\ead{jafranck@wisc.edu}


\address[uw-affiliation]{Department of Engineering Physics, College of Engineering, University of Wisconsin-Madison, Madison, WI, USA - 53706}

\address[engg-affiliation]{School of Engineering, Brown University, Providence, RI, USA - 02912}





\begin{abstract}

Bio-inspired oscillatory foil propulsion has the ability to traverse various propulsive modes by dynamically changing the foil's heave and pitch kinematics. This research characterizes the propulsion properties and wake dynamics of a symmetric oscillating foil, specifically targeting the high Reynolds number operation of small to medium surface vessels whose propulsive specifications have a broad range of loads and speeds. An unsteady Reynolds-averaged Navier-Stokes (URANS) solver with a k-$\omega$ SST turbulence model is used to sweep through pitch amplitude and frequency at two heave amplitudes of $h_0/c=1$ and $h_0/c=2$ at $Re=10^6$. At $h_0/c=2$, the maximum thrust coefficient is $C_T=8.2$ due to the large intercepted flow area of the foil, whereas at a decreased Strouhal number the thrust coefficient decreases and the maximum propulsive efficiency reaches 75\%. Results illustrate the kinematics required to transition between the high-efficiency and high-thrust regimes at high Reynolds number and the resulting changes to the vortex wake structure. The unsteady vortex dynamics throughout the heave--pitch cycle strongly influence the characterization of thrust and propulsive efficiency, and are classified into flow regimes based on performance and vortex structure.

Declarations of interest: none

\end{abstract}


\end{frontmatter}


\section{Introduction} \label{sec:intro}

Oscillating foil propulsion (OFP) is inspired from thunniform swimming in fish and other marine organisms, and offers an alternative propulsion strategy to rotary-based propellers. An oscillatory heaving and pitching foil can be either drag-producing or thrust-producing depending on its kinematic parameters, namely frequency, heave and pitch amplitudes, and phase difference between heave and pitch. Various kinematics have been well documented to correspond to high propulsive efficiency regimes, and other regimes of high thrust. A major advantage to OFP is the ability to dynamically change the foil's heave and pitch kinematics and thus naturally traverse various propulsive modes. Drawing inspiration from aquatic animals, the same device can operate in a high-thrust, high-maneuverability regime as an animal would in an escape mode, and then transition to a high-efficiency or cruise-mode by modifying its flapping frequency and/or amplitude. Due to its oscillatory, rather than rotational motion, OFP offers other advantages such as quieter operation and less adverse effects on marine life due to lower tip speeds. 

\begin{figure}[h]
  \centering
  \includegraphics[width=0.5\textwidth]{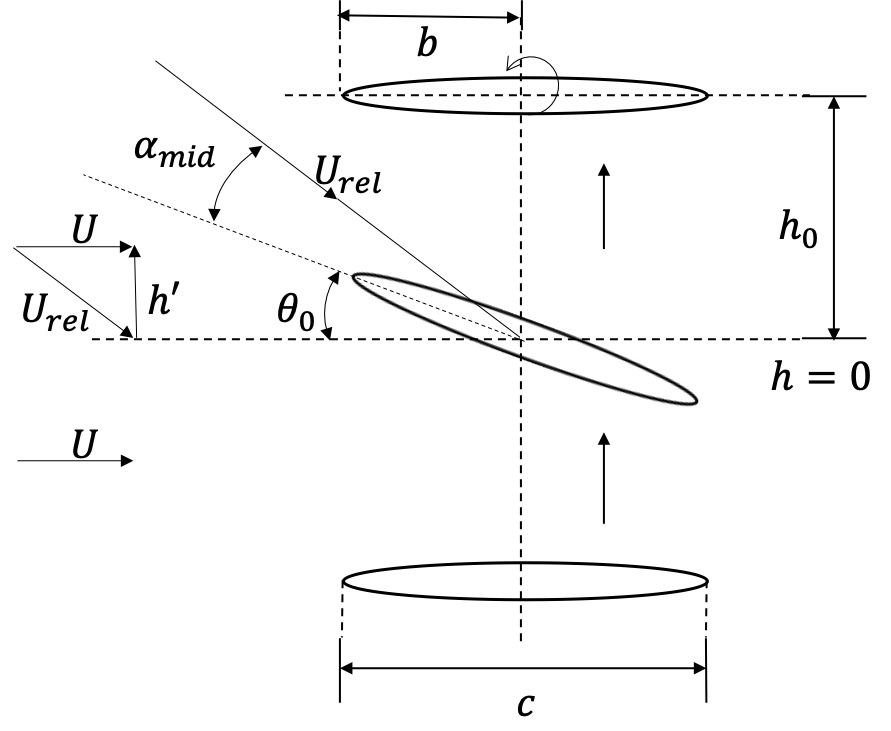}
  \caption{A schematic diagram demonstrating various dimensions and kinematic parameters of the foil motion.}
  \label{fig:nomencl}
\end{figure}

Previous research on oscillating foils is well summarized in review articles by \cite{Rozhdestvensky2003} and \cite{Triantafyllou2004a}. Figure \ref{fig:nomencl} shows a schematic diagram of a foil's oscillatory motion, defined by a heaving amplitude $h_0$, pitching amplitude $\theta_0$, and the pivot location $b$. Most commonly, the flapping frequency $f$ is non-dimensionalized in terms of the Strouhal Number $St$, 

\begin{equation}
\label{eqn:St}
    St = \frac{2 h_0 f}{U},
\end{equation}
and is a function of the heave amplitude and freestream velocity $U$. Previous research has shown that the Strouhal number is the principal parameter governing thrust generation and wake dynamics \citep{Triantafyllou1991,Ramamurti2001}.  The ideal operating range in terms of thrust and efficiency has been found to be between $St=0.2$ to $0.4$, with pitch amplitudes between $40^{\circ}$ and $60^{\circ}$, which has been demonstrated experimentally \citep{Fish1998, Anderson1998, Read2003, Hover2004, Schouveiler2005, Techet2007} and computationally \citep{Tuncer1996, Jones1997, Young2004, Xiao2010, LaMantia2011, Mattheijssens2012}. The effect of heave amplitude has received less attention, but typical values range from $h_0/c=0.5$ to $1$. Larger heave amplitudes of $h_0/c>1$ have been performed by \cite{Isogai1999} and \cite{Katz1978} finding that the thrust increases with heave amplitude and that the efficiency is highest in regimes where no flow separation occurs. Using a potential flow model, \cite{Floch2012} also explored higher heave amplitudes and directly compared to conventional propellers by defining an equivalent advance parameter for oscillating foils. Change in efficiency with the advance parameter (frequency) was characterized for different heave and pitch amplitudes equivalent to the propeller efficiency curves for different relative pitch values.

Typically, the oscillatory motion is prescribed via a sinusoidal heave and pitch motion separated by a phase angle of $\phi=90^{\circ}$. Small variations in $\phi$ yielded no significant changes to thrust and efficiency \citep{Read2003}, however modifying the sinusoidal trajectory has shown modest improvements. The effective angle of attack, given by

\begin{equation}
\label{eqn:alpha}
    \alpha(t) = \theta(t) - tan^{-1} \left( \frac{h'(t)}{U} \right)
\end{equation}
for a sinusoidal stroke (neglecting the effects from angular velocity), can exhibit multiple maxima/minima in each upstroke/downstroke for high pitch and heave amplitudes.  Thus, controlling the shape of $\alpha(t)$ directly has demonstrated increases up to 50\% in maximum thrust \citep{Read2003, Hover2004} and the degradation of thrust force and efficiency at high Strouhal number was found to be alleviated \citep{Xiao2010}). Investigations have shown that with careful control, chordwise or spanwise flexibility can also improve the performance of OFP \citep{Katz1978, Liu1997}. In particular, \cite{Richards2015} found a correlation between the frequency ratio (frequency of oscillation to resonant frequency) of a flexible foil and its propulsive efficiency.  However the efficiency gains of flexibility in OFP carry the trade-off of added cost and complexity of materials and a control system. 

Various propulsive modes such as high-efficiency regimes or high-thrust regimes, demonstrate distinct characteristics in the wake. Early experiments by \cite{Triantafyllou1991} and \cite{KOOCHESFAHANI2008} documented the vortical flow patterns and the presence of a jet profile, or reverse von K\'arm\'an street in the wake. \cite{Lai1999} visually demonstrated the wake vortices changing from drag producing to thrust producing while increasing the heave amplitude or frequency of a plunging foil. More recently, \cite{Andersen2017} conducted a combined numerical and experimental study on wake structures with pure heaving compared with pure pitching, and \cite{Liu2017} analyzed the wake structure and performance for low aspect ratio flapping foils by running three-dimensional flow simulations. 
A computational investigation by \cite{Zurman-Nasution2020} at a Reynolds number of $Re=5.3 \times 10^3$ showed that the flow structures and performance of a heaving foil in propulsive mode are determined by two-dimensional effects at an intermediate Strouhal number of around $St \approx 0.3$, whereas three-dimensional effects dominate at lower and higher $St$. These results were observed to hold for a $Re \approx 10^4$, however the nature of three-dimensionality at higher Reynolds numbers has not been explored.

Most of the computational and experimental work described above has been performed at low to moderate Reynolds numbers $(Re<10^5)$ targeting the propulsion properties of aquatic animals, or the design of highly maneuverable OFP for small unmanned underwater vehicles (UUVs). OFP is also an attractive propulsion strategy for small to medium surface vessels that have diverse operating conditions and a wide range of loads such as tugs, fishing vessels, and wind turbine repair vessels. OFP offers more than just propulsion including inherent stability and enhanced maneuverability with multiple foils, which can eliminate the need for other costly control systems. Experiments and simulations at high Reynolds number $(Re>10^5)$ have been limited likely due to laboratory limitations on measuring and testing as well as the effects of turbulence which can be challenging to model and computationally intensive. 

Those that have explored higher Reynolds numbers include \cite{Isogai1999}, who used a compressible Navier-Stokes solver to simulate flow around a pitching--heaving airfoil. Different phase angles and reduced frequencies were performed with laminar and turbulent simulations using the Baldwin and Lomax method at $Re=10^5$. Within the kinematics regimes investigated, the laminar and turbulent flow simulation results were found to be almost identical. \cite{Ashraf2011} used a Reynolds-averaged Navier-Stokes (RANS) model to look at the effect of foil thickness at a Reynolds number of $Re=10^3$ up to $2 \times 10^6$. It was found that thin airfoils were favorable at low Reynolds number but thicker airfoils were favorable at high Reynolds number. 
Although these studies provide valuable insight into performance of oscillating foils in high Reynolds number turbulent flows, they only encompass a narrow parameter range, and give little attention to the turbulent wake structure. 

The objective of this paper is to document performance and wake structure of OFP at $Re = 10^6$, a Reynolds number regime that has received little attention in academic studies, particularly for kinematics at high heave amplitudes. These results not only fill a niche in the scientific literature in terms of kinematics and Reynolds number, but also provide relevant baseline performance metrics for the marine propulsion industry in terms of OFP design. By sweeping a large range of 126 unique kinematics, the 2D RANS simulations presented in this paper give a broad overview of the propulsive regimes, associated wake structures, and propulsive capabilities. 
Simulations are performed with a 10\% thick elliptic foil with a pitching motion about mid-chord ($b=c/2$). The fore-aft symmetric foil offers simplicity to design and includes industrial applications that may benefit from fully functional forward and reverse operations. Simulations are performed at $h_0/c=1$ and $h_0/c=2$ at seven Strouhal numbers and nine pitch amplitudes for each heave amplitude. The range of kinematics thoroughly covers the thrust generation regime in which the mid-stroke angle of attack, $\alpha_{mid}$, is in the range of interest $\numrange{10}{25}^\circ$. The cumulative efficiency and thrust performance of the high Reynolds number and high-heave kinematics are analyzed and the resulting vortex wake dynamics are described and classified based on performance and wake structure modes.

\section{Computational Setup} \label{sec:methods}

\subsection{Numerical methods}
The simulations solve the unsteady Reynolds-averaged Navier-Stokes (URANS) equations for incompressible flow given by, 

\begin{equation}
    {\nabla} \cdot \ubar = 0
    \label{eqn:rans_cont}
\end{equation}

\begin{equation}
    \frac{\partial \ubar}{\partial t} + (\ubar \cdot {\nabla}) \ubar = 
        - {\nabla} \overline{p} + \nu {\nabla}^2 \ubar - {\nabla} \cdot \bm{\tau}
    \label{eqn:rans_mom}
\end{equation}
where $\ubar$ is the mean component of the velocity vector and $\overline{p}$ is the mean pressure. The Reynolds stress tensor $\bm{\tau}$ in equation \ref{eqn:rans_mom} is modeled with the $k$-$\omega$ SST equations \citep{Menter1994}. This turbulence model is chosen due to the likelihood of separated flow, but it is also compared against three other turbulence models in Section \ref{subsec:turb_model} to examine model sensitivity.

\begin{figure}[h]
  \centering
  \includegraphics[width=0.5\textwidth]{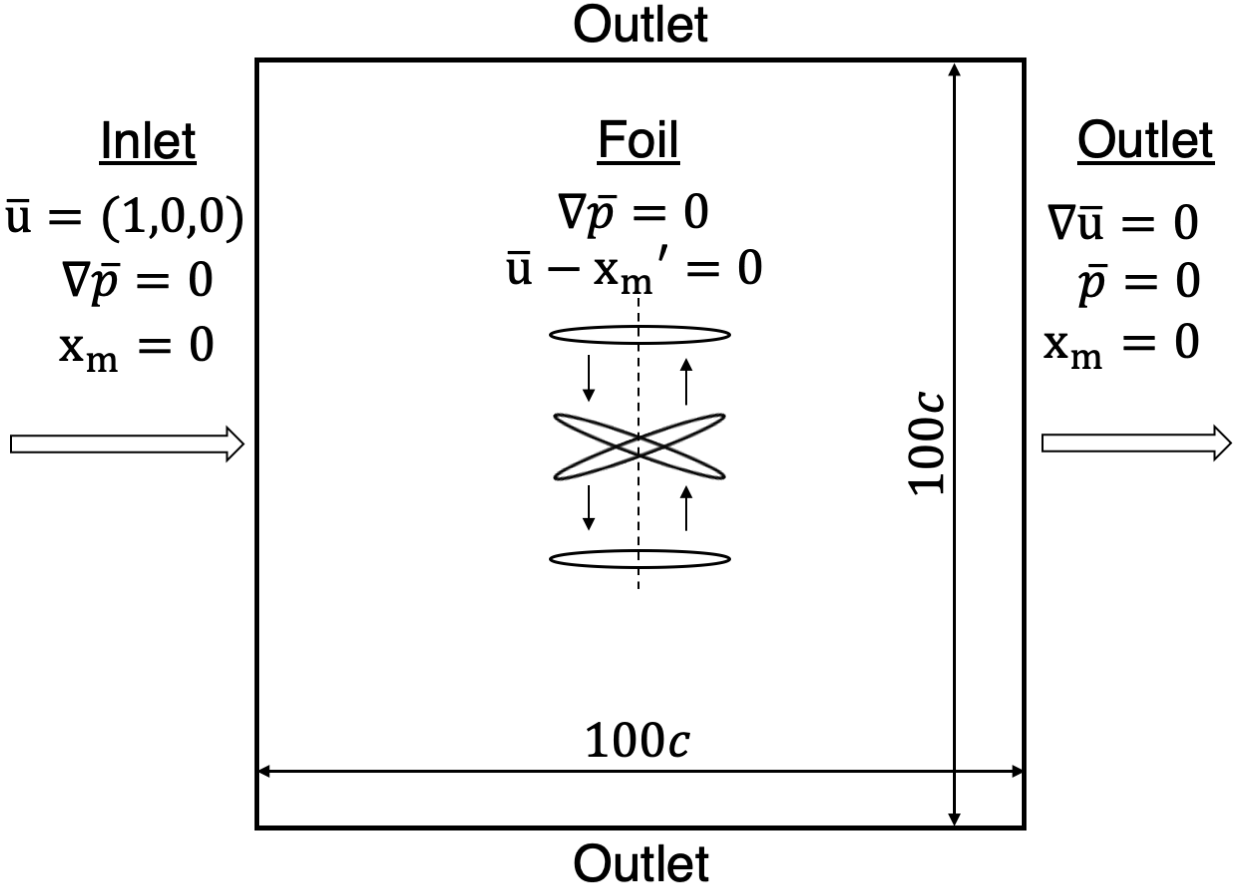}
    \caption{Boundary conditions at the inlet, outlet and foil surface.}
  \label{fig:BC}
\end{figure}

A second-order accurate finite volume, pressure-implicit split-operator (PISO) method \citep{Issa1986} is implemented using OpenFOAM \citep{Weller1998}. The solver implements a first-order accurate implicit (Euler) time integration scheme. The pressure corrector step is solved with a geometric--algebraic multi-grid (GAMG) algorithm for three iterations each time-step with a tolerance of $10^{-6}$. 
As shown in figure \ref{fig:BC} the computational domain contains inlet and outlet boundary conditions, and is $100$ chord lengths in both the streamwise and transverse directions.  

To account for motion of the foil, a dynamic meshing routine is implemented such that the displacement of mesh elements, $\mathbf{x_m}$, is computed at each time iteration according to the equation

\begin{equation} \label{eqn:dynmesh}
    2 \nabla \cdot \left[ \mu \nabla \mathbf{x_m} \right] + {\nabla} \cdot \left[ \mu \left( ({\nabla} \mathbf{x_m})^T  - {\nabla} \mathbf{x_m} - \mathbf{I}\ tr({\nabla} \mathbf{x_m}) \right) \right] = 0
\end{equation}
which is solved with a conjugate gradient method. The mesh motion equation contains a diffusivity constant $\mu$ which determines how nodal displacement will be distributed among the surrounding cells as the foil moves. For the simulations described here, $\mu$ varies as the inverse of distance from the foil. This provides minimal mesh deformation and skewness in close proximity to the foil, and the large computational domain provides a large area for the deformations to take place in the far-field.  The displacement at the outer boundary is set to zero, such that the overall size of the domain remains fixed. An unstructured mesh is utilized as it is found to be more resilient to the mesh motion and deformations than a structured mesh.  The current mesh has approximately 80,000 cells and is developed with the software Gmsh \citep{Geuzaine2009}.  Three different zones of mesh resolution are used, with the resolution increasing closer to the foil surface.  At the foil's surface in the undeformed mesh, the resolution at $x/c=0.5$ is approximately $\Delta x/c = 0.0003$. Wall functions are utilized to calculate the value of turbulent kinetic energy, $k$, and the specific dissipation, $\omega$, for the first layer of mesh cells at the foil based on distance from the wall \citep{Liu2016}. A zero-gradient boundary condition is imposed for $k$, and a blended function of the viscous and log-law variations from \cite{Menter2001} is used for $\omega$, providing the flexibility for the first mesh layer to be in the viscous sub-layer or the log-law region.

\begin{figure}
  \centering
    \begin{subfigure}{.3\textwidth}
        \centering
        \includegraphics[trim=300 220 300 220, clip, width=\textwidth]{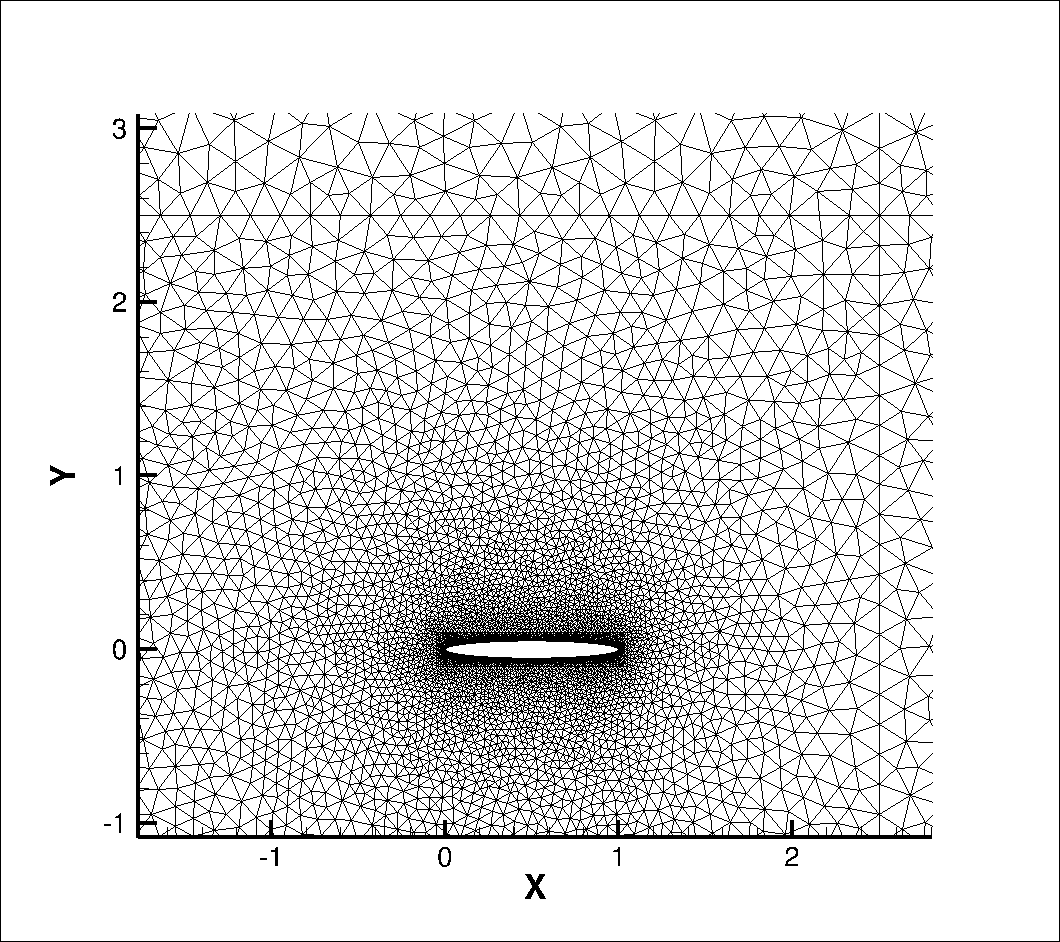}
        \caption{Bottom of stroke\\ $(t/T \approx 0)$}
    \end{subfigure}
    \begin{subfigure}{.3\textwidth}
        \includegraphics[trim=300 220 300 220, clip, width=\textwidth]{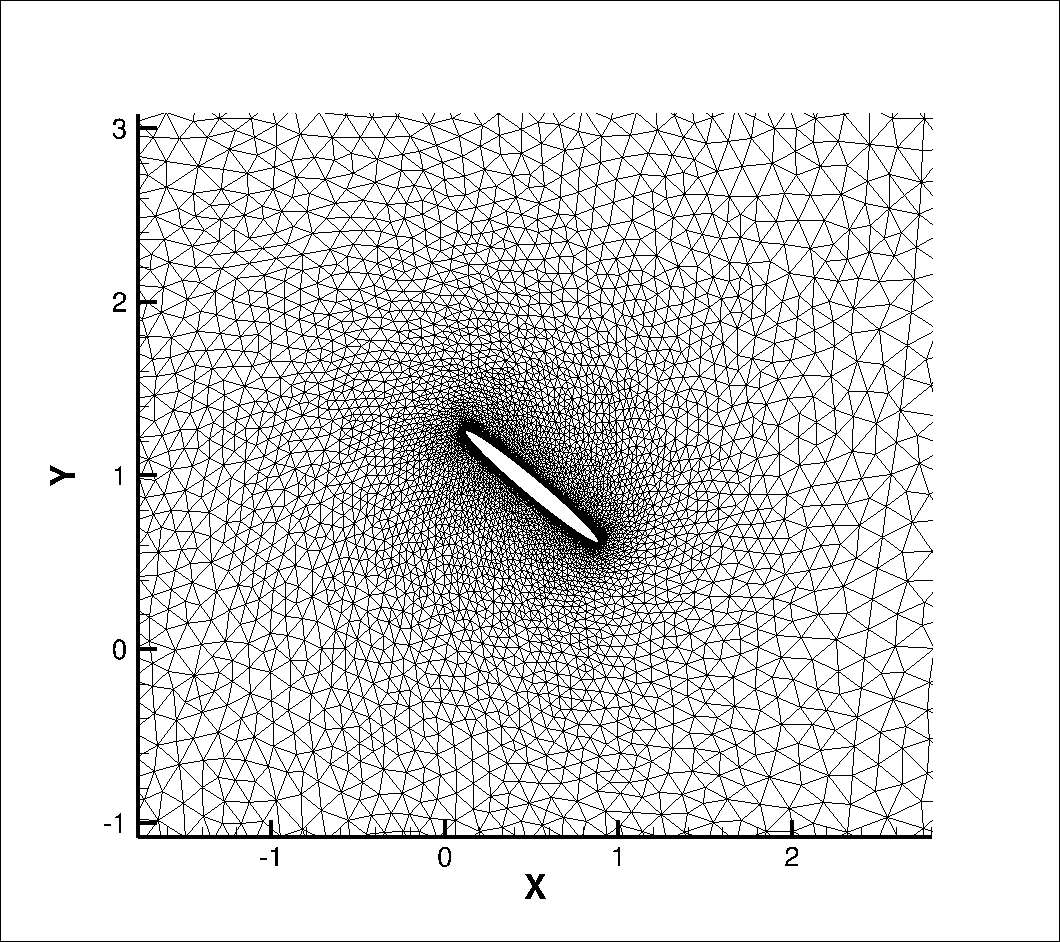}
        \caption{Mid-upstroke\\ $(t/T \approx 0.25)$}
        \label{subfig:dynmesh_h_0}
    \end{subfigure}
    \begin{subfigure}{.3\linewidth}
        \includegraphics[trim=300 220 300 220, clip, width=\textwidth]{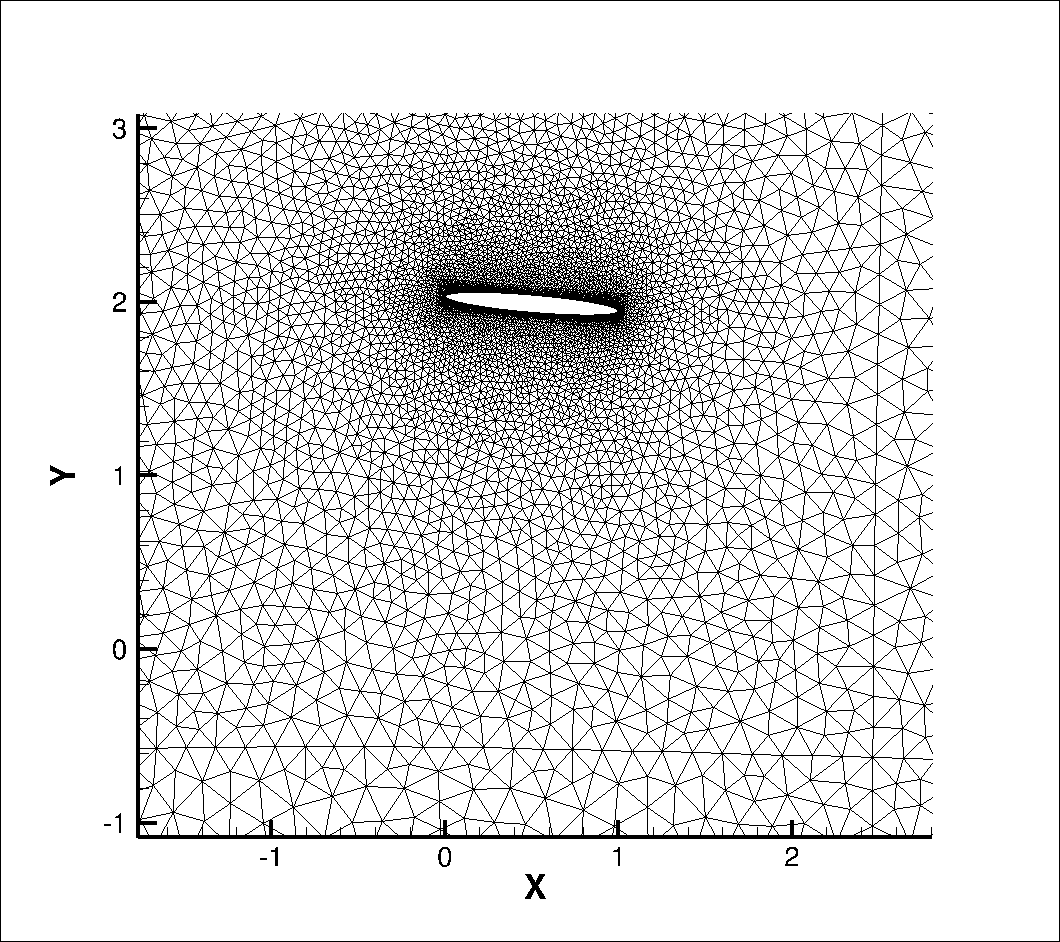}
        \caption{Top of stroke\\ $(t/T \approx 0.5)$}
        \label{subfig:dynmesh_h_h0}
    \end{subfigure}
    \caption{The dynamic mesh at three different positions during an upstroke ($ h_0/c = 1.0, \theta_0 = 40^\circ, St = 0.533 $).}
  \label{fig:dynmesh}
\end{figure}

Figure \ref{fig:dynmesh} demonstrates the dynamic meshing at three different positions during a typical upstroke with $h_0/c=1$. The rotation of the foil imposes a skewness in the mesh elements at a radial line approximately one chord from the center of the foil in figure \ref{subfig:dynmesh_h_0}, but then the mesh rotates back to its original position at the top of the stroke.  As the foil heaves in the upward direction, the mesh elements above the foil become slightly more concentrated while those below the foil are stretched.  However this is accounted for in the original mesh element distribution such that the mesh resolution retains its symmetry very closely.

\subsection{Performance metrics}

Oscillating foil simulations are performed at a Reynolds number, $Re=Uc/\nu$, of $10^6$.
The simulations are divided into two sets indicated by heave amplitudes $h_0/c=1$ and $h_0/c=2$, respectively. For each heave amplitude performance metrics are reported at seven different Strouhal numbers and nine different pitch amplitudes, or 63 distinct kinematics per heave amplitude. For the purposes of validation an additional set of simulations are performed at $h_0/c=0.75$ with $b=c/3$, to compare with closely correlated experimental data \citep{Schouveiler2005}.

\begin{figure}
  \centering
  \includegraphics[width=0.9\textwidth]{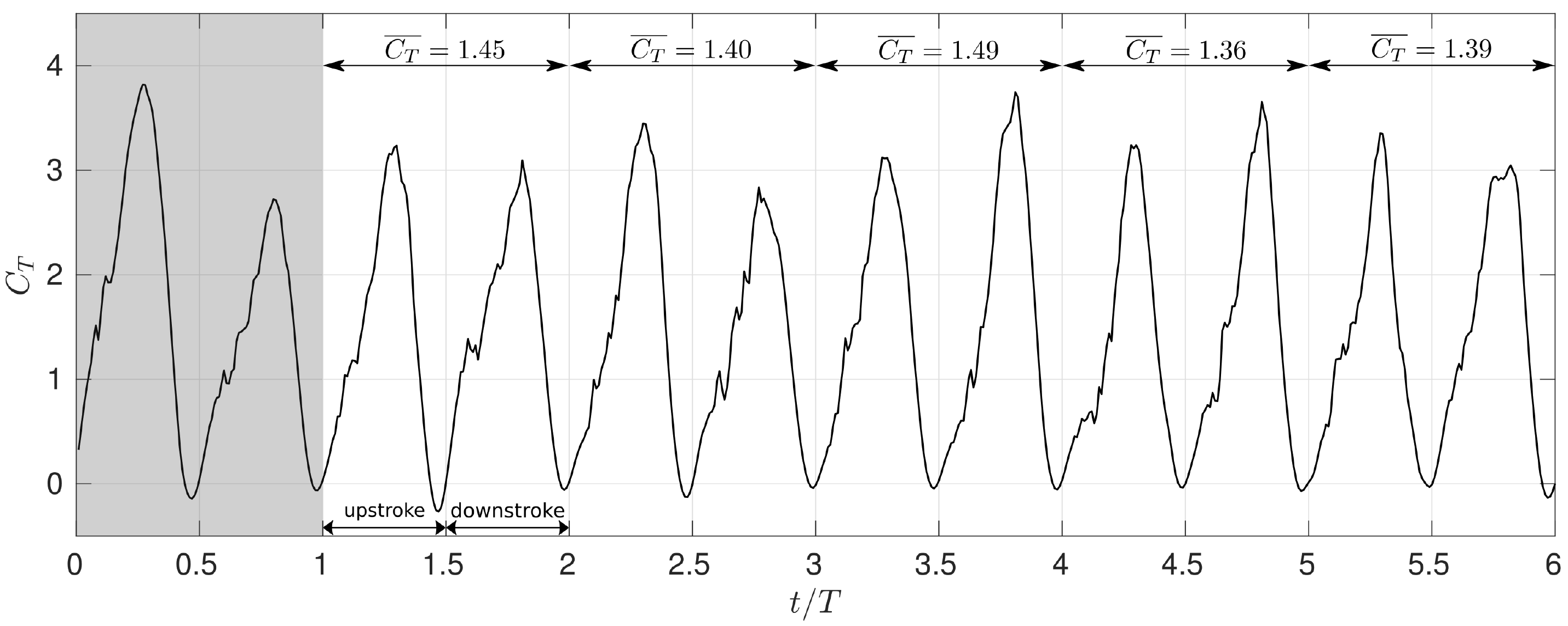}
    \caption{Thrust coefficient profile over six cycles of oscillation, for the kinematics: $ h_0/c = 1.0, \theta_0 = 45^\circ, St = 0.6 $, shows minor variations between each period. The first cycle is not included in the mean thrust coefficient.}
\label{fig:ct_all_periods}
\end{figure}

Each simulation is run in parallel with 16 MPI processes on the Oscar compute cluster at Brown University. A typical compute node consists of 24 cores with Intel's Haswell architecture. The IBM General Parallel File System (GPFS) is utilized for storage and the nodes are connected via a 40 Gigabit per second Infiniband network. Shell scripting and job arrays are used to automate the process of running the simulations and post-processing the results.  As an initial condition, a steady state boundary layer is allowed to develop by running the flow simulation on a static foil with zero angle of attack.  Then each computation is simulated for six oscillation cycles, taking 24 to 48 hours to complete depending on the prescribed frequency of oscillation. Figure \ref{fig:ct_all_periods} demonstrates the thrust coefficient profile for a pitching/heaving foil simulation.  Due to the strong inertial force of the pitching and heaving kinematics, it takes as little as one heaving cycle to reach its fully developed propulsive state, although there are minor variations between each period. For the propulsive metrics, the mean value over the last five cycles is reported.

Performance of the oscillating foil is measured in terms of the horizontal thrust force generated ($F_x$), or the force exerted on the foil opposite to the flow velocity. It is non-dimensionalized in form of the thrust coefficient, 

\begin{equation}
\label{eqn:CT}
    C_T(t) = \frac{F_x(t)}{\frac{1}{2} \rho U^2 c},
\end{equation}
where $\rho$ is density of the fluid. Of importance to the efficiency is also the input power, $P(t)$, required to move the foil. Power required for the heaving motion is calculated as the product of vertical force on the foil $F_y(t)$ and the heaving velocity $h'(t)$. Power required for the pitching motion is a product of span-wise moment on the foil $M_z(t)$ and the pitching velocity $\theta'(t)$. The input power is non-dimensionalized as 

\begin{equation}
\label{eqn:powercoeff}
    C_P(t) = - \frac{F_y(t) h'(t) + M_z(t) \theta'(t)}{\frac{1}{2} \rho U^3 c},
\end{equation}
where the negative sign is required as power needs to be input when the force or moment is acting against the velocity. The net propulsive efficiency of the system, $\eta$, is a ratio of the mean thrust coefficient to the mean input power coefficient over each cycle, or

\begin{equation}
\label{eqn:efficiency}
    \eta = \overline{C_T} / \overline{C_P}.
\end{equation}

As defined in equation \ref{eqn:CT} the thrust coefficient is non-dimensionalized by planform area of the foil. Unlike a rotational propeller the intercepted area of an oscillating foil will change with its prescribed kinematics, sweeping a distance of approximately $2h_0$. With an increase in swept area, one can expect a larger thrust due to the additional momentum transfer, and an alternative thrust coefficient metric provided by \cite{Floch2012} is defined by

\begin{equation}
\label{eqn:CTstar}
    C_T^*(t) = \frac{F_x(t)}{\frac{1}{2} \rho U^2 A},
\end{equation}
where $A=2h_0$ in a two-dimensional simulation.

\subsection{Verification and validation of model}
\label{subsec:turb_model}

\begin{figure}
  \centering
  \includegraphics[width=0.9\textwidth]{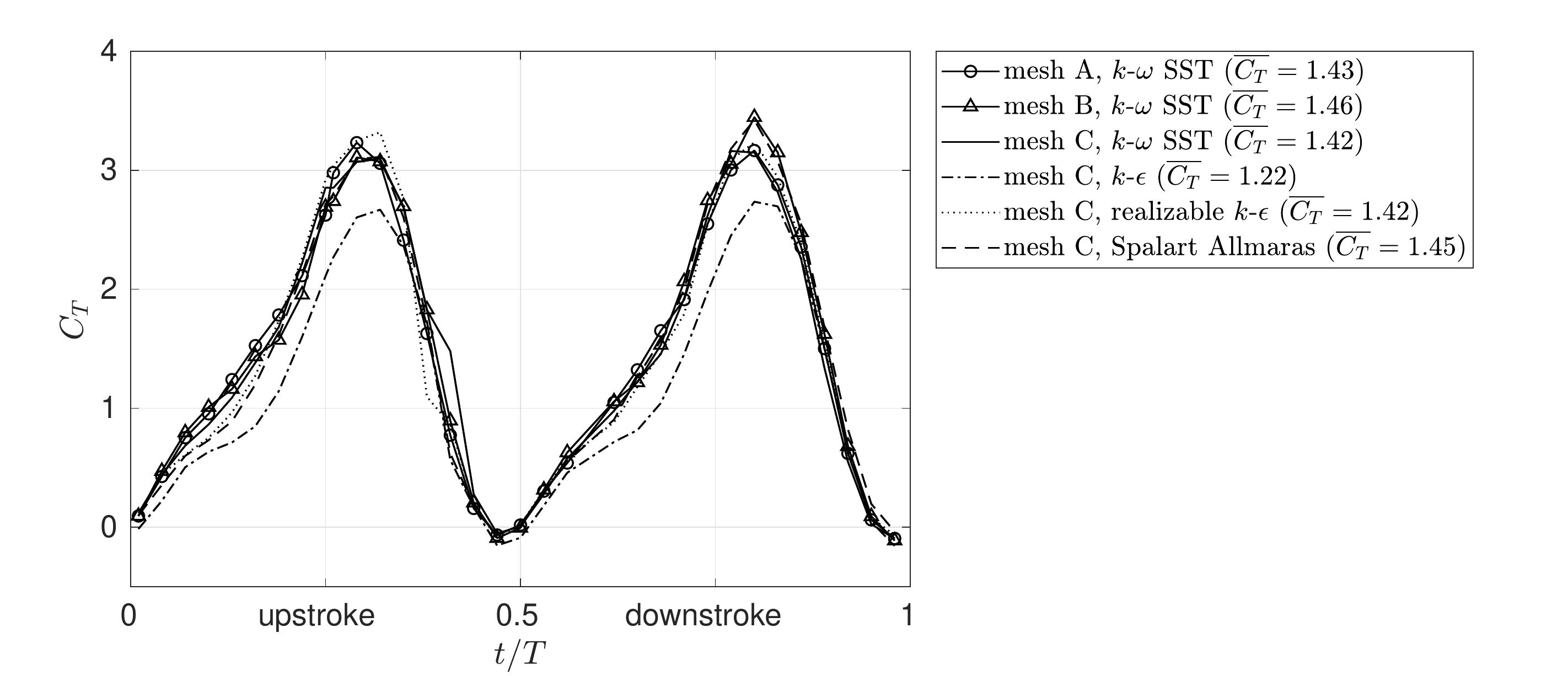}
    \caption{Comparison of phase-averaged thrust coefficient for four different RANS closure models and three different mesh resolutions, for the kinematics: $ h_0/c = 1.0, \theta_0 = 45^\circ, St = 0.6 $.}
\label{fig:ct_turb}
\end{figure}

In order to assess the mesh resolution three different meshes are compared in table \ref{tab:meshes}. 
For a stationary foil in uniform flow at zero angle of attack, the maximum value of dimensionless wall distance ($y^+$) for the first layer of mesh cells in mesh C is $1.87$ while the average $y^+$ is $0.1$. Further reducing the resolution at the foil does not provide a stable solution. The three meshes are compared directly for a single set of kinematics ($ h_0/c = 1.0, \theta_0 = 45^\circ, St = 0.6$) and the phase-averaged, or time-dependent thrust forces averaged over five cycles are shown in figure \ref{fig:ct_turb}. Given the little variation between the three meshes, mesh C is considered adequate resolution for the simulations.

\begin{table}
  \begin{center}
    \caption{Performance metrics for the kinematics: $ h_0/c = 1.0, \theta_0 = 45^\circ, St = 0.6 $ with different mesh configurations using the $k$-$\omega$ SST turbulence model. The maximum and average $y^+$ values at the first mesh layer are for a stationary foil in uniform flow at zero angle of attack.}
    \begin{tabularx}{0.97\textwidth}{C D C C D C C} \toprule
      Mesh identifier & $\Delta x / c$ for first mesh layer & $y^+_\textrm{max}$ & $y^+_\textrm{avg}$ & Total number of cells & $\overline{{C_T}}$ & $\overline{{C_P}}$ \\
      \midrule
      \textbf{mesh A} & 0.0006 & 2.90 & 0.43 & 50,581 & 1.43 & 2.79 \\
      \textbf{mesh B} & 0.0003 & 1.54 & 0.05 & 159,508 & 1.46 & 2.83 \\
      \textbf{mesh C} & 0.0003 & 1.87 & 0.10 & 79,986 & 1.42 & 2.75  \\
      \bottomrule
    \label{tab:meshes}
    \end{tabularx}
  \end{center}
\end{table}

\begin{figure}
  \centering

    \begin{subfigure}{.4\textwidth}
        \includegraphics[width=\textwidth]{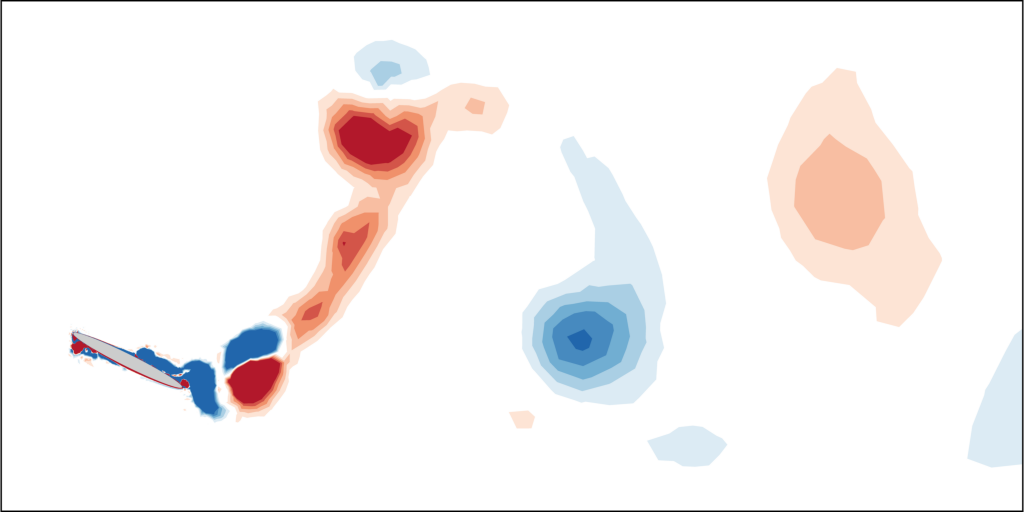}
        \caption{mesh C}
    \end{subfigure}\hspace{.5 cm}
    \begin{subfigure}{.4\textwidth}
        \includegraphics[width=\textwidth]{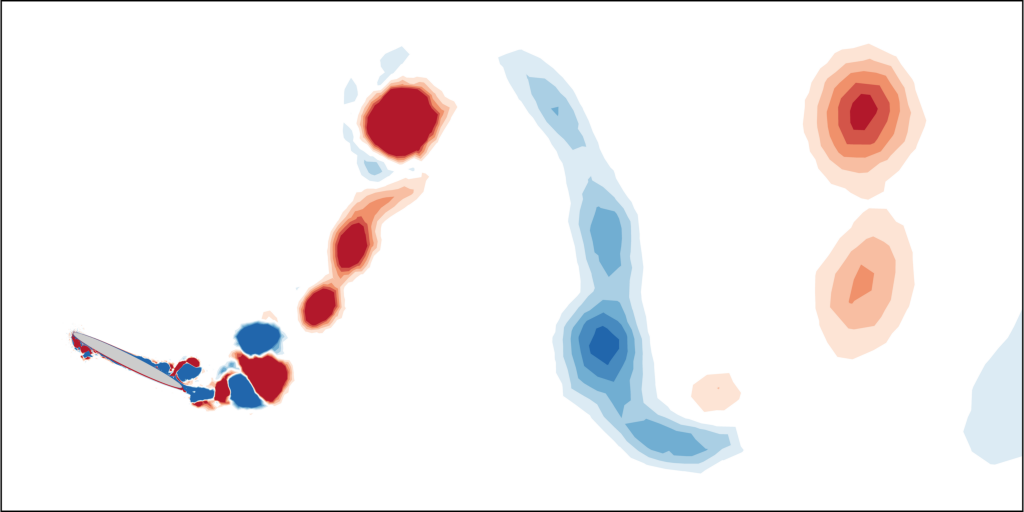}
        \caption{mesh B}
    \end{subfigure}
    
  \caption{Vorticity contours demonstrating the effect of mesh resolution in the wake, for kinematics $ h_0/c = 1.0, \theta_0 = 45^\circ, St = 0.6 $.}
  \label{fig:wake_res}
\end{figure}

In figure \ref{fig:wake_res}, contours of vorticity are compared between mesh C and mesh B so as to highlight the effect of resolution in the wake.
The coarser resolution in mesh C results in over-dissipation of the vortices, but the overall structure of the wake can be observed from the vortices shedding at the foil. The analysis presented in section \ref{subsec:hydrodynamics} requires a broad classification of the wake structure into different regimes. Hence mesh C is chosen for the simulations to reduce the turn-around time.

Figure \ref{fig:ct_turb} also compares simulations for the same kinematic parameters using four different turbulence models, including the $k$-$\omega$ SST model that is ultimately chosen, for determining the sensitivity to the choice of turbulence model. All of the turbulence models, (realizable $k$-$\epsilon$, Spalart Allmaras and $k$-$\omega$ SST), gave similar quantitative and qualitative results with the exception of the $k$-$\epsilon$ model which consistently had lower thrust.
The $k$-$\epsilon$ model had high turbulent viscosity values and performed poorly in predicting boundary layer separation. 
The model however mandates the use of a wall function for the dissipation rate, $\epsilon$, that requires the first cell layer to be in the log-law region. Although mesh C is too resolved for this model, it is challenging to create a mesh that fulfills this requirement throughout the periodic motion of the foil. The realizable $k$-$\epsilon$ variant of the model however is able to predict the performance close to the other two models, likely due to the dynamic computation of a coefficient in the transport equations.
Ultimately the $k$-$\omega$ SST model is chosen due to its documented ability to handle separated flows \citep{Bardina1997}.

\begin{figure}
  \centering
    \includegraphics[width=.8\linewidth]{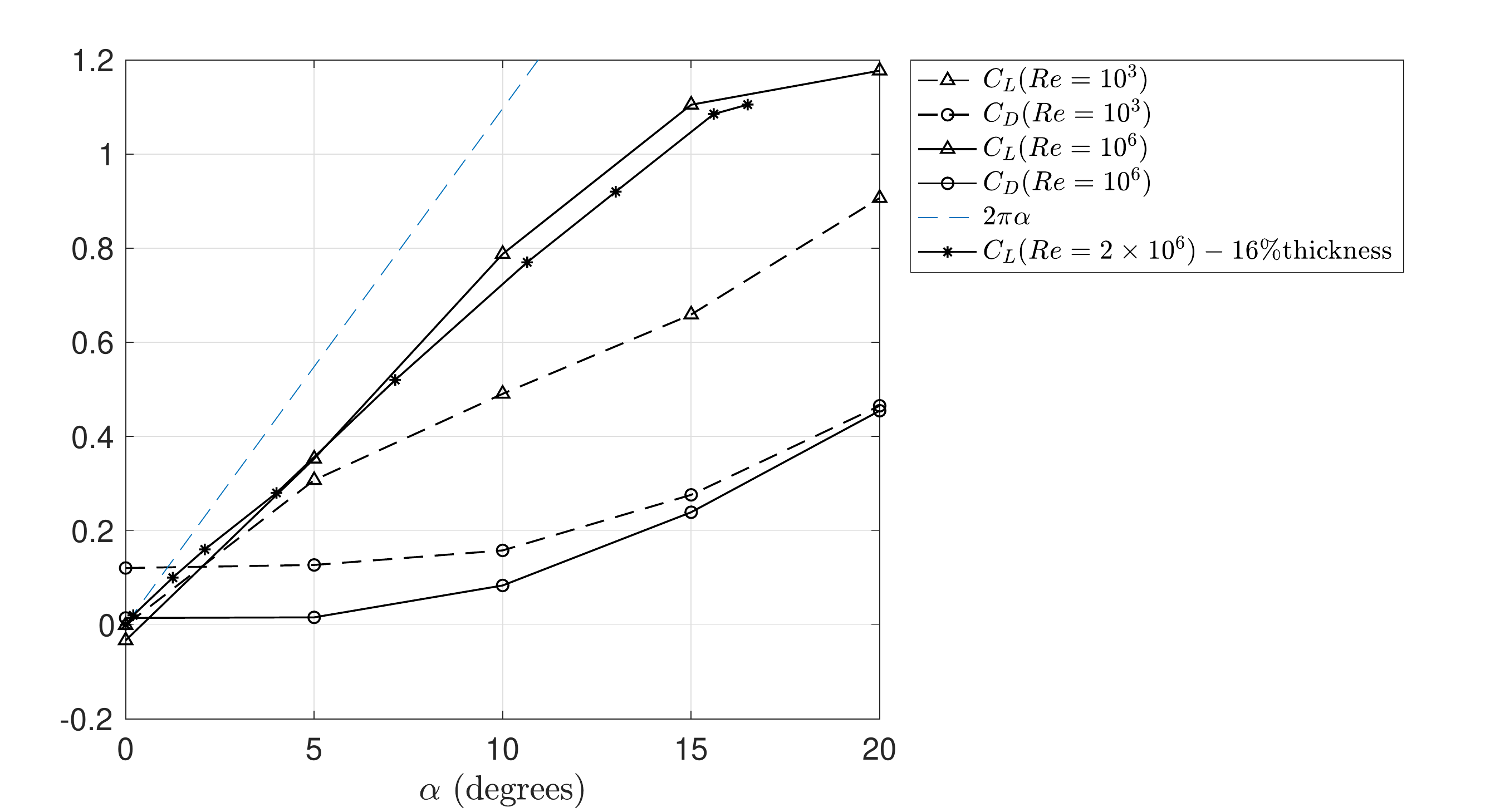}
      \caption{Lift and drag coefficients on a stationary 10\% thick ellipse as a function of angle of attack for DNS at $Re = 10^3$ (dashed lines) and RANS at $Re = 10^6$ (solid lines) show significant Reynolds number effects. Results compared against available experimental $C_L$ data of an elliptic foil with 16\% thickness \citep{Hoerner1985}.}
\label{fig:comp-polars}
\end{figure}

Lift and drag curves for a stationary 10\% thick ellipse at low and high Reynolds number are computed in figure \ref{fig:comp-polars}. Experimental data of lift coefficient at $Re=2 \times 10^6$ for an elliptic foil with 16\% thickness \citep{Hoerner1985} is included for comparison and shows good agreement with the computed lift coefficient at $Re=10^6$ for the 10\% thick elliptic foil.

\section{Results and Discussion} \label{sec:results}

\subsection{Performance of an elliptical foil and effects of Reynolds number}
\label{subsec:compareRe}

\begin{figure}
  \centering
  \begin{subfigure}{.4\textwidth}
  \centering
  \includegraphics[width=\linewidth]{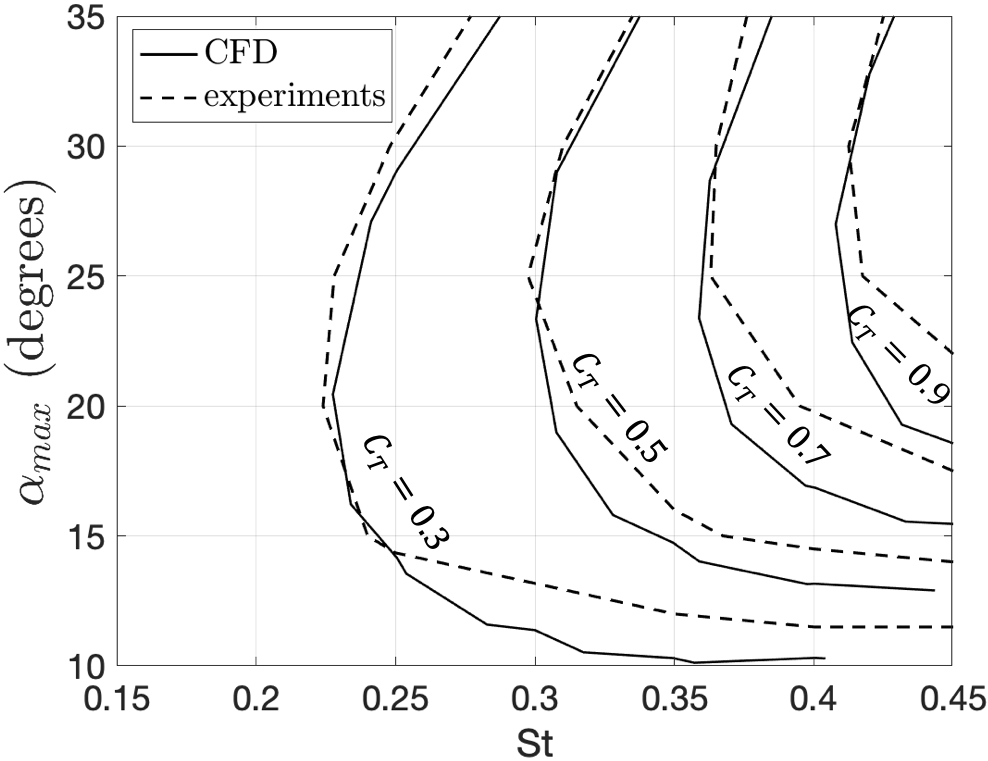}
  \caption{$C_T$ (thrust coefficient)}
  \label{fig:validationa}
  \end{subfigure} \hspace{0.36cm}
\begin{subfigure}{.4\textwidth}
  \centering
  \includegraphics[width=\linewidth]{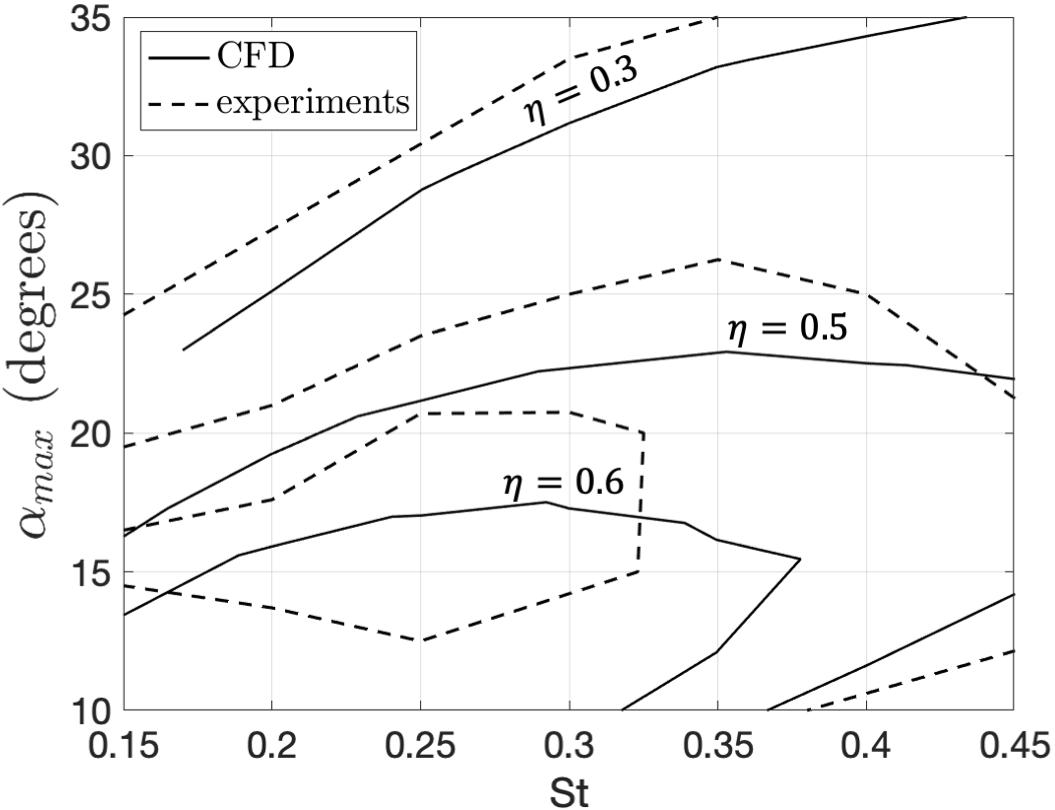}
  \caption{$\eta$ (efficiency)}
    \label{fig:validationb}
\end{subfigure}
    \caption{Comparison of contour lines of efficiency and thrust coefficient with results from \cite{Schouveiler2005}.}
\label{fig:validation}
\end{figure}

Simulations at $h_0/c=0.75$ and pivot location $b=c/3$ are performed to compare with similar experimental data reported by \cite{Schouveiler2005}. Although the kinematics are well matched, the simulations contain an elliptic foil at $Re=10^6$ whereas a NACA 0012 foil at a lower Reynolds number of $Re=4\times10^4$ was utilized in the experiments. Figure \ref{fig:validation} compares the performance over the range of Strouhal numbers and maximum relative angle of attack, $\alpha_{max}$, where $\alpha$ is given by equation \ref{eqn:alpha}. Despite the differences in foil shape and Reynolds number, very good agreement for thrust coefficient is demonstrated in figure \ref{fig:validationa}. The propulsive efficiency in figure \ref{fig:validationb} has good qualitative agreement, documenting the same trends for changes in $\alpha_{max}$ and $St$, with the experiments achieving higher efficiency. The differences in efficiency between simulation and experiment are more dramatic at low relative angles of attack when the boundary layer is fully attached, which is likely due to the improved lift and drag coefficients of the NACA 0012 compared to an elliptic foil.

\begin{figure}[t!]
  \centering
  
  \begin{subfigure}{0.8\textwidth}
  \centering
  \includegraphics[width=0.45\linewidth]{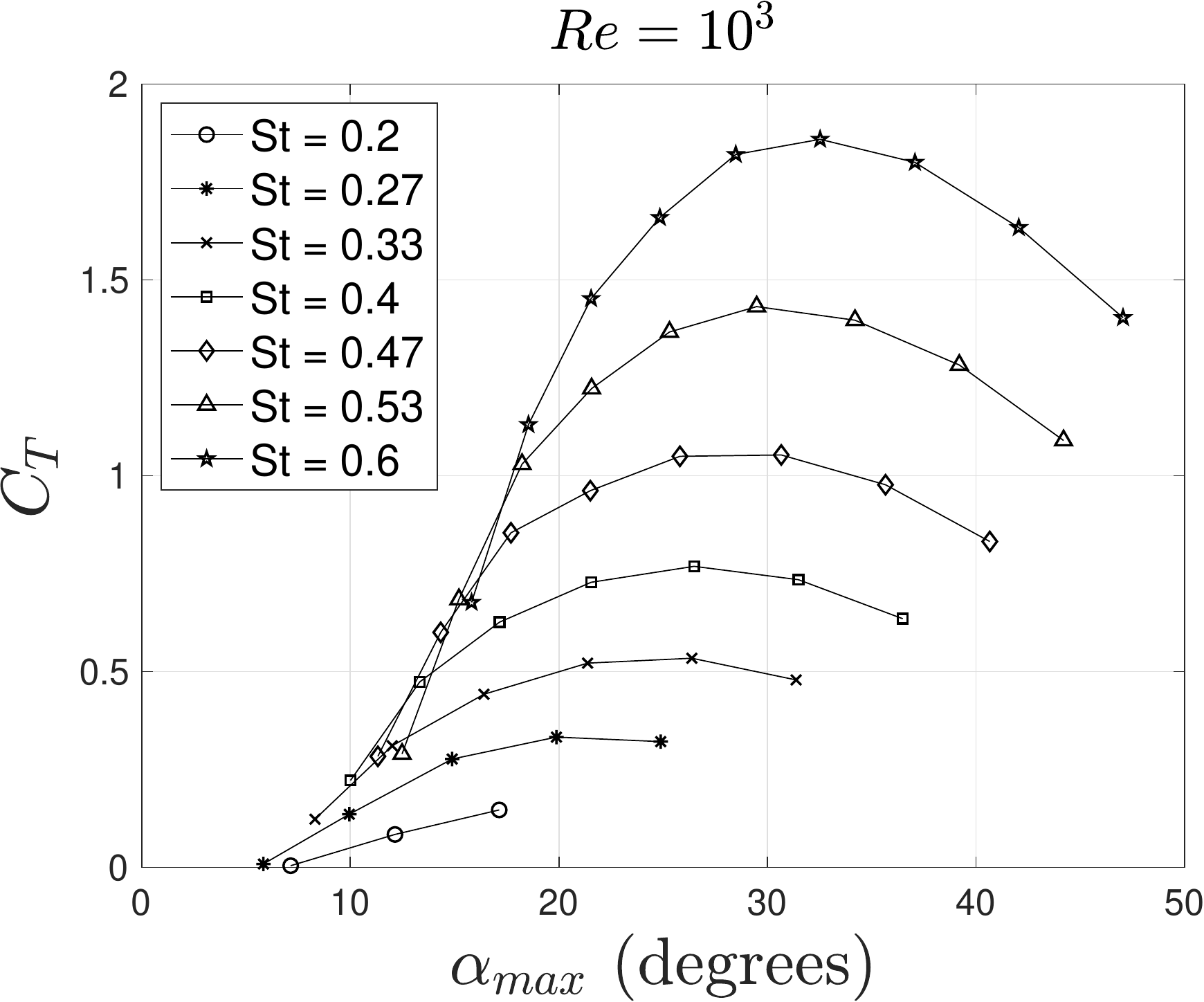} \hspace{.5 cm}
  \includegraphics[width=0.45\linewidth]{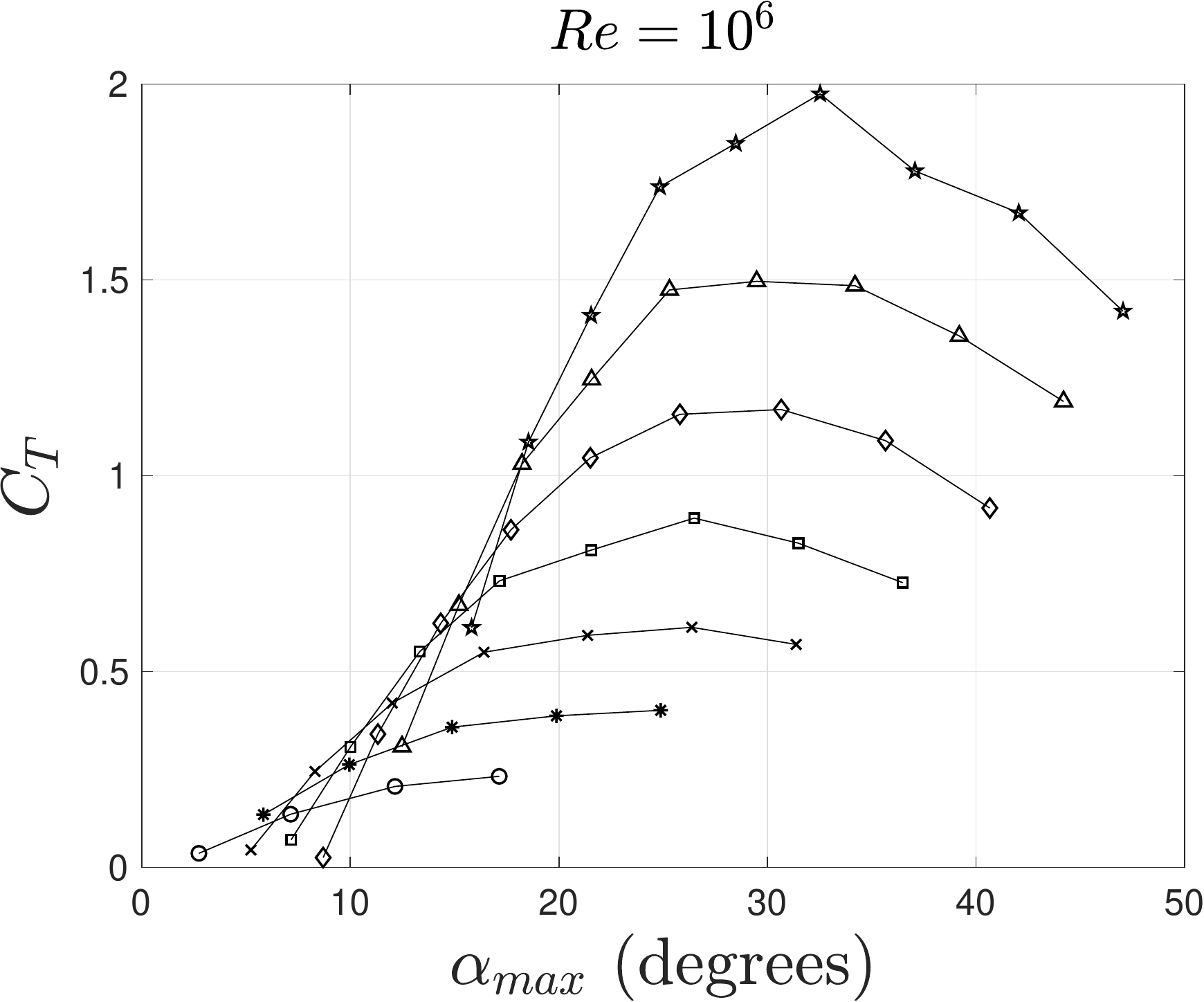}
  \caption{$C_T$ (thrust coefficient)}
  \end{subfigure}
  
  \begin{subfigure}{0.8\textwidth}
  \centering
  \includegraphics[width=0.45\linewidth]{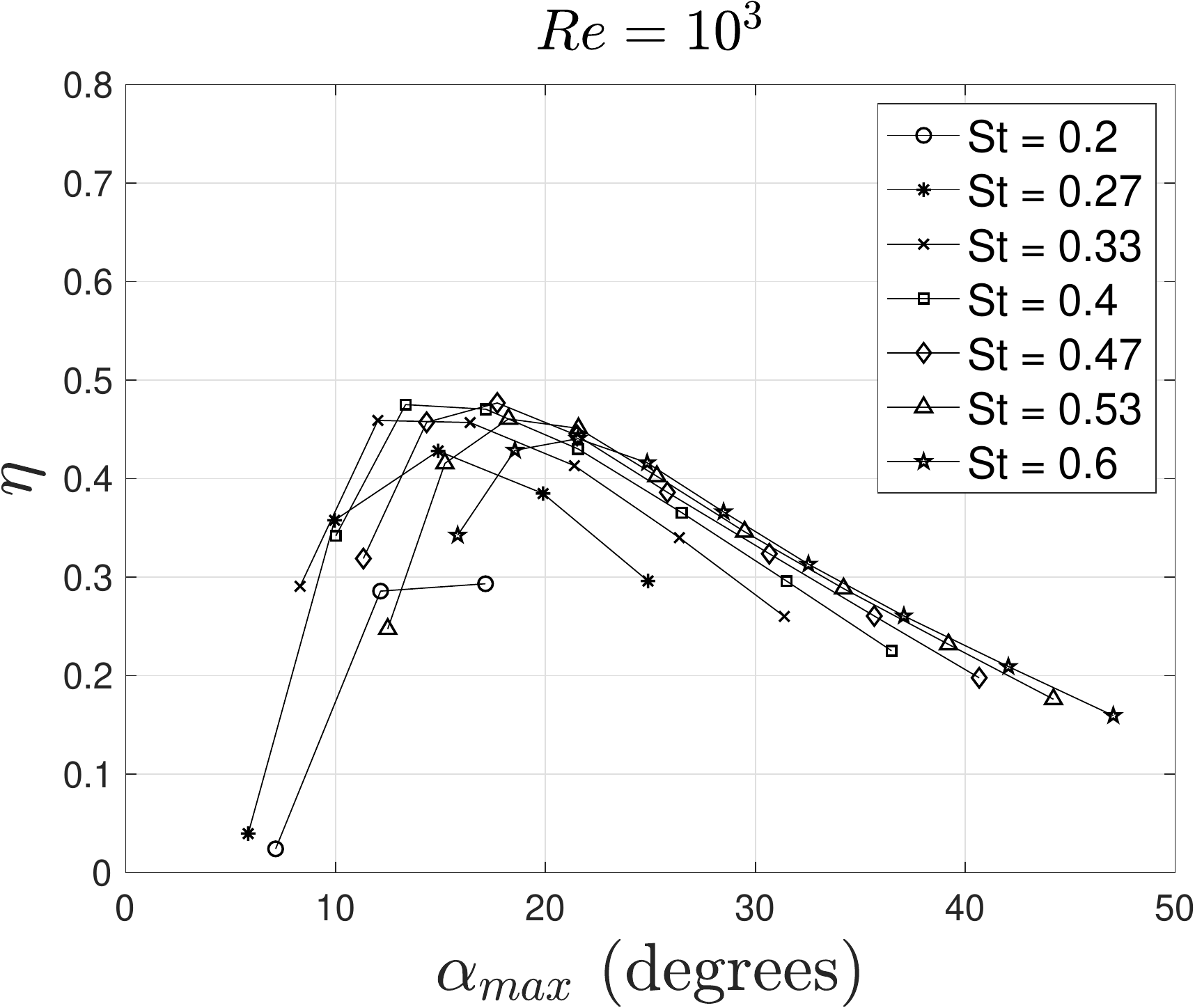} \hspace{.5 cm}
  \includegraphics[width=0.45\linewidth]{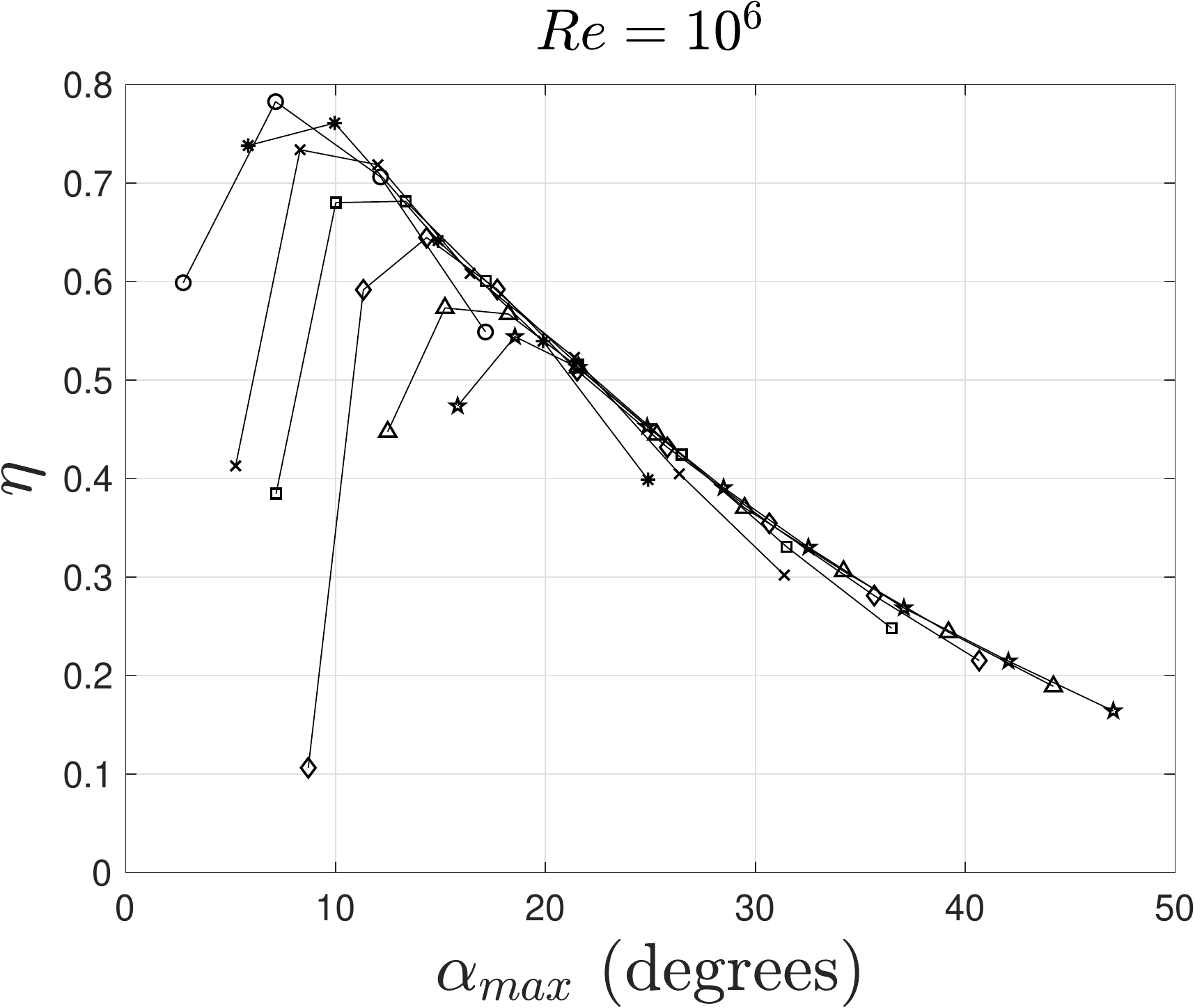}
  \caption{$\eta$ (efficiency)}
  \end{subfigure}
    \caption{Comparison of DNS at $Re=10^3$ (left) and RANS simulations at $Re=10^6$ (right).}

  \label{fig:comp-laminar}
\end{figure}

To isolate the effect of Reynolds number, RANS simulations at $Re=10^6$ are compared with DNS results at $Re=10^3$ for the simulations with $ h_0/c = 1 $. The efficiency and thrust coefficient as a function of $\alpha_{max}$ for discrete Strouhal numbers are shown in figure \ref{fig:comp-laminar}. Overall, the two Reynolds numbers demonstrate similar trends by increasing thrust with increasing Strouhal number, with the maximum occurring at approximately $\alpha_{max}=30^{\circ}$. In both Reynolds numbers the efficiency drops off at high $\alpha_{max}$ but has different trends at low $\alpha_{max}$. The peak efficiency values are higher for RANS simulations than for the low Reynolds number DNS. At low Reynolds number the efficiency peaks around \numrange{40}{50}\% for all Strouhal numbers. At high Reynolds number, the peak efficiency increases significantly with decreasing Strouhal number and is at its highest at 78\% for $St=0.2$. All efficiency values greater than 60\% occur at low $\alpha_{max}$, or fully attached flow regimes. Within this regime, the thrust coefficients are improved with increased Reynolds number, which in turn affects the efficiency, causing a drastic difference between Reynolds numbers for $0^{\circ}<\alpha_{max}<20^{\circ}$. At these angles of attack there is very little thrust produced. However the RANS also shows greater efficiency compared to the DNS within the regime of $15^{\circ}<\alpha_{max}<30^{\circ}$, which has significantly greater thrust than the fully attached (and highly efficient) low angle of attack regime. 

\begin{figure}[t!]
  \centering
  \includegraphics[width=0.8\linewidth]{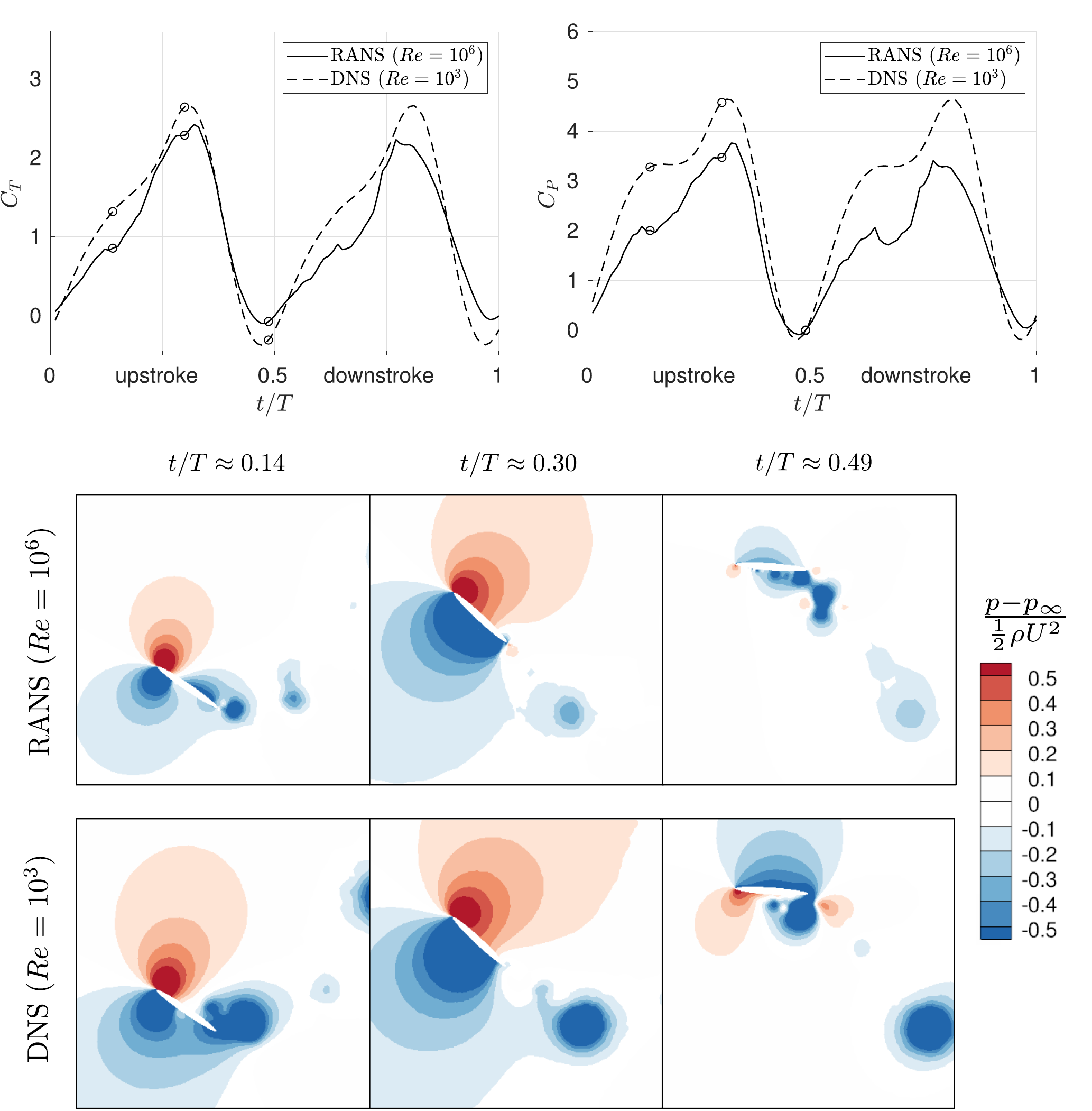}
    \caption{Comparison of performance (top) and pressure (bottom) for the kinematics: $ h_0/c = 1.0, \theta_0 = 45^\circ, St = 0.534 $ at $Re = 10^3$ and $Re = 10^6$ shows how high Reynolds number results in higher efficiency.}
    \label{fig:comp-all}
\end{figure}

The above differences in Reynolds number can be explained by a combination of two related factors. 
As observed in figure \ref{fig:comp-polars}, there is a strong Reynolds number dependence for the ellipse foil shape, which has also been noted by \cite{Kwon2005}.
Secondly, OFP is significantly affected by flow separation and the resulting formation of a leading edge vortex (LEV). With increasing Reynolds number, a turbulent boundary layer is more resistant to separation and thus these vortex dynamics are modified or delayed.
A detailed comparison is shown for a specific set of kinematics ($ h_0/c = 1.0, \theta_0 = 45^\circ, St = 0.534 $) in figure \ref{fig:comp-all} with the variation of phase-averaged thrust and input power coefficients, along with contours of normalized pressure at three different positions during the oscillation cycle. The maximum angle of attack encountered for these kinematics is $\alpha_{max} = 18^\circ$ at which a turbulent boundary layer will be more resistant to separation than the laminar boundary layer. Hence a stronger LEV is created at $Re=10^3$ because of more dramatic flow separation, as seen most clearly at $t/T \approx 0.30$, which results in higher peak thrust values.
However, due to a much lower lift-to-drag ratio as observed in figure \ref{fig:comp-polars}, the net-force vector at $Re=10^3$ has a much higher vertical component than at $Re=10^6$ resulting in consistently higher input power required. Additionally, when the foil is at zero angle of attack at the end of each stroke, the drag is higher at $Re=10^3$ which results in attrition of the average thrust generated. The above effects cumulatively result in an average thrust and efficiency of $C_T = 1.18, \eta=45.9\%$ at $Re=10^3$ and $C_T = 1.03, \eta=56.7\%$ at $Re=10^6$.

\subsection{Wake structure and effect of hydrodynamics} \label{subsec:hydrodynamics}

Because of the highly dynamic nature of these simulations, different kinematics result in different flow regimes and wake characteristics, which have been analyzed closely in previous studies \citep{KOOCHESFAHANI2008, Lai1999, Hover2004, Andersen2017}. The categorization of vortex patterns in the wake of an oscillating cylinder by \cite{Williamson1988} has been applied to oscillating foils using the format ``$mS+nP$" to denote vortex shedding patterns, where $m$ is the number of individual or single ({\it S}) vortices shed during each oscillation cycle and $n$ is the number of vortex pairs ({\it P}). It is difficult to discern such distinct vortex patterns at high Strouhal numbers and high angles of attack due to the multiple degrees of freedom and the turbulence in the wake at high Reynolds number. Moreover, it is not known how accurately the two-dimensional model represents the physical flow-field, as the flow structures have been found to have three-dimensional effects in certain kinematic regimes for an oscillating foil at lower Reynolds numbers of $O(10^4)$ \citep{Zurman-Nasution2020}. However, the main shedding structure and vortex organization is strongly dependent on kinematics and thus categorizations can be applied based on the performance metrics and observed vortex patterns.

\begin{figure}[b]
  \centering
    \includegraphics[height=0.75cm]{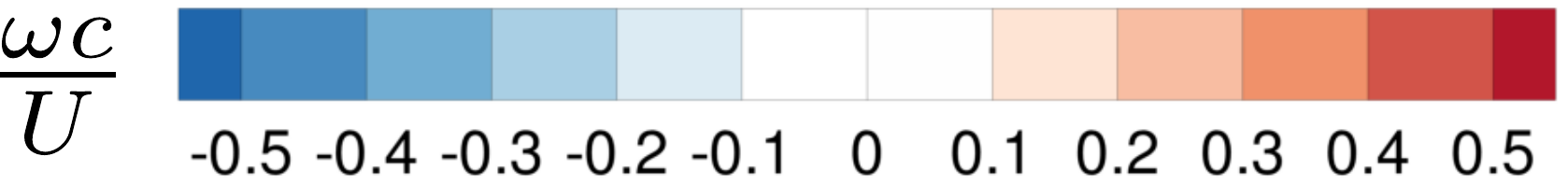}

    \begin{subfigure}{.4\textwidth}
        \includegraphics[width=\textwidth]{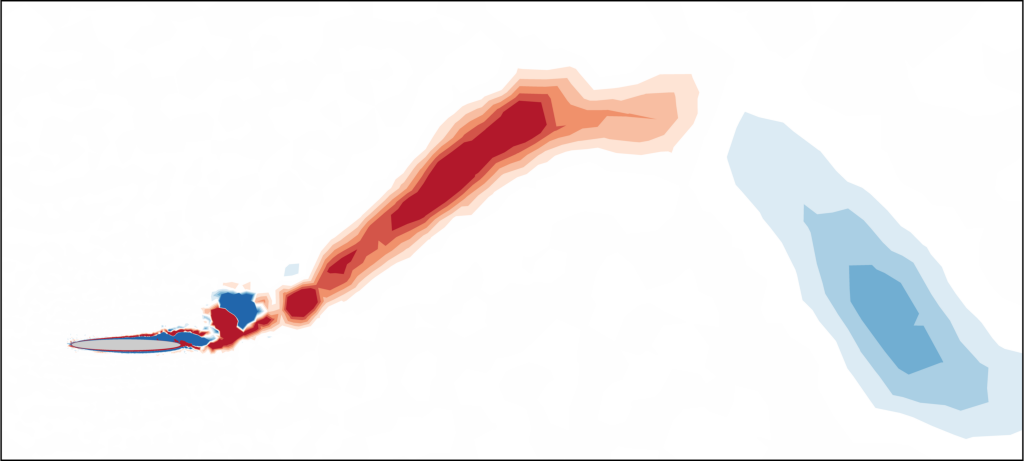}
        \caption{$ t/T \approx 0 $}
    \end{subfigure}\hspace{.5 cm}
    \begin{subfigure}{.4\textwidth}
        \includegraphics[width=\textwidth]{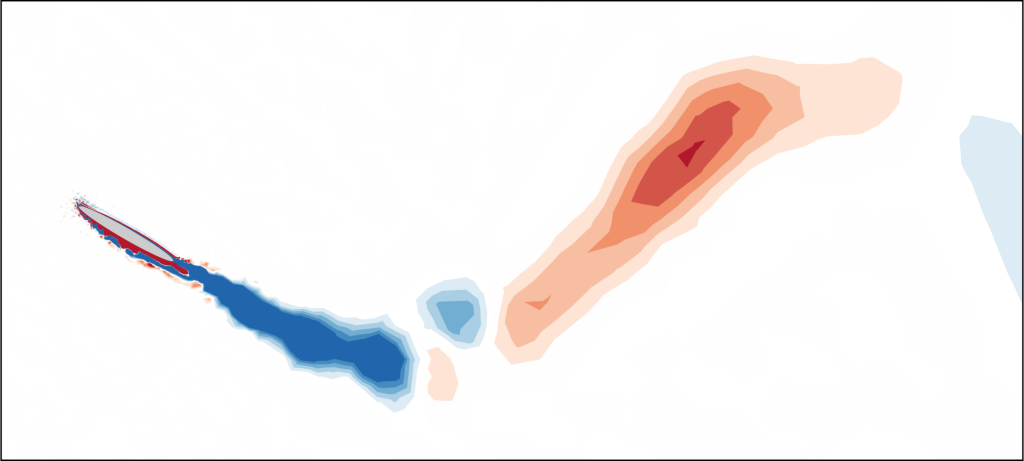}
        \caption{$ t/T \approx 0.25 $}
    \end{subfigure}
  \caption{Regime A vorticity contours demonstrated with kinematics $ h_0/c = 1.0, \theta_0 = 30^\circ, St = 0.267 $.}
  \label{fig:regimeA}
\end{figure}

\begin{figure}
  \centering
    \includegraphics[height=0.75cm]{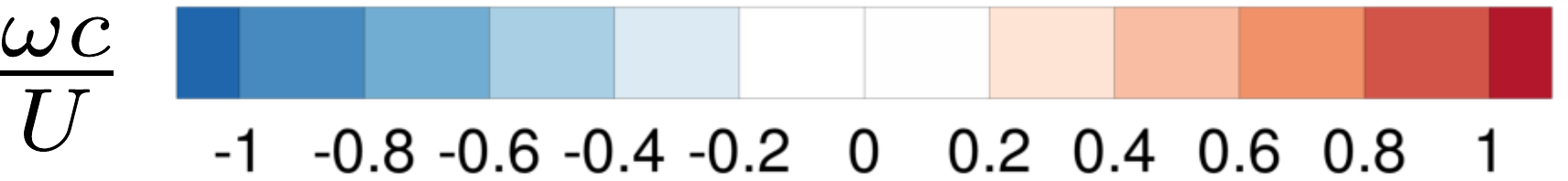}

    \begin{subfigure}{.4\textwidth}
        \includegraphics[width=\textwidth]{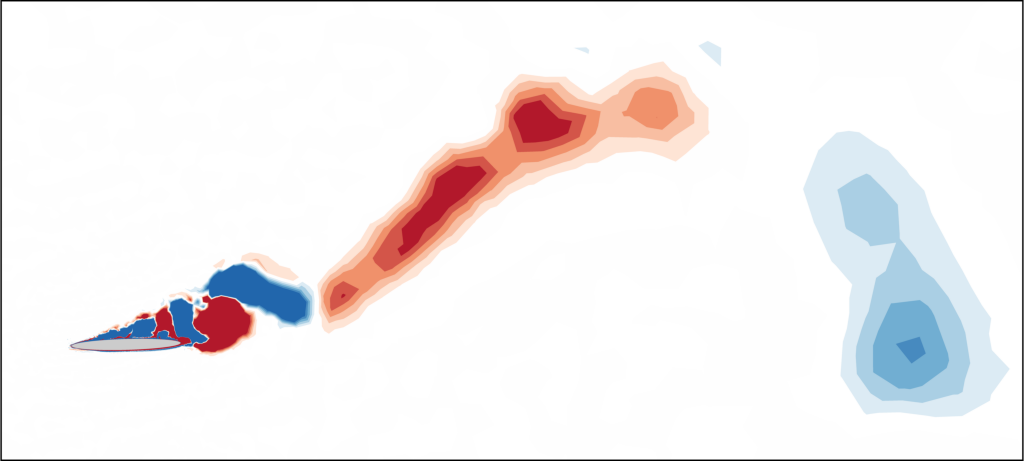}
        \caption{$ t/T \approx 0 $}
    \end{subfigure}\hspace{.5 cm}
    \begin{subfigure}{.4\textwidth}
        \includegraphics[width=\textwidth]{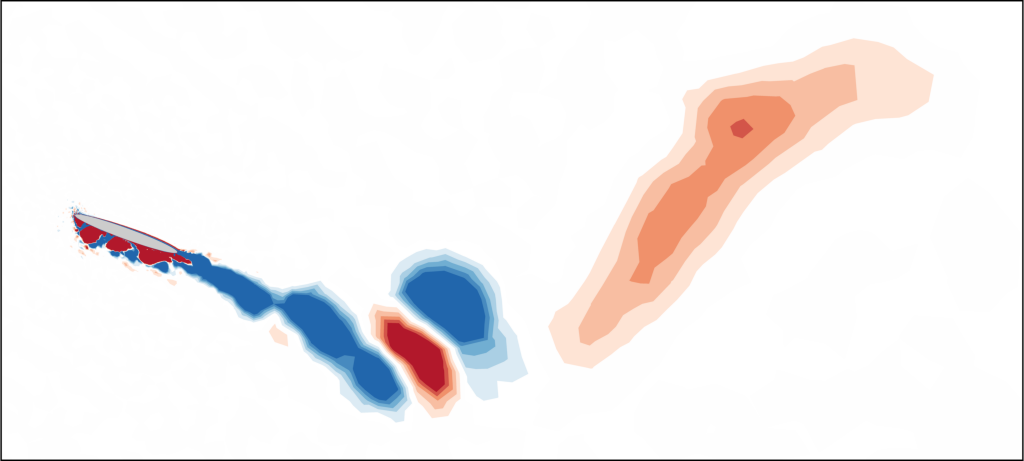}
        \caption{$ t/T \approx 0.25 $}
    \end{subfigure}
  \caption{Regime B vorticity contours demonstrated with kinematics $ h_0/c = 1.0, \theta_0 = 20^\circ, St = 0.267 $.}
  \label{fig:regimeB}
\end{figure}

\begin{figure}[t]
  \centering
    \includegraphics[height=0.75cm]{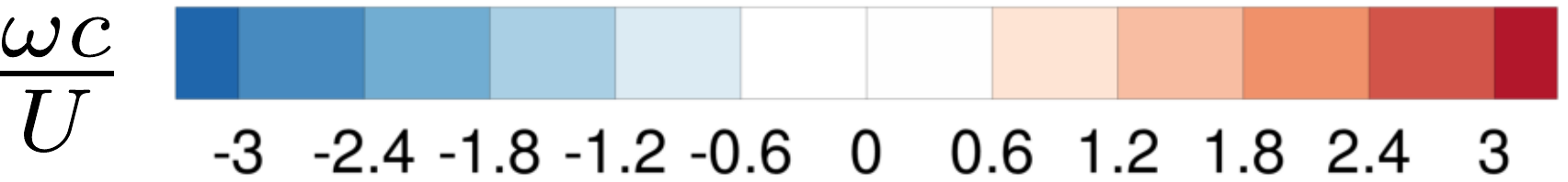}

    \begin{subfigure}{.4\textwidth}
        \includegraphics[width=\textwidth]{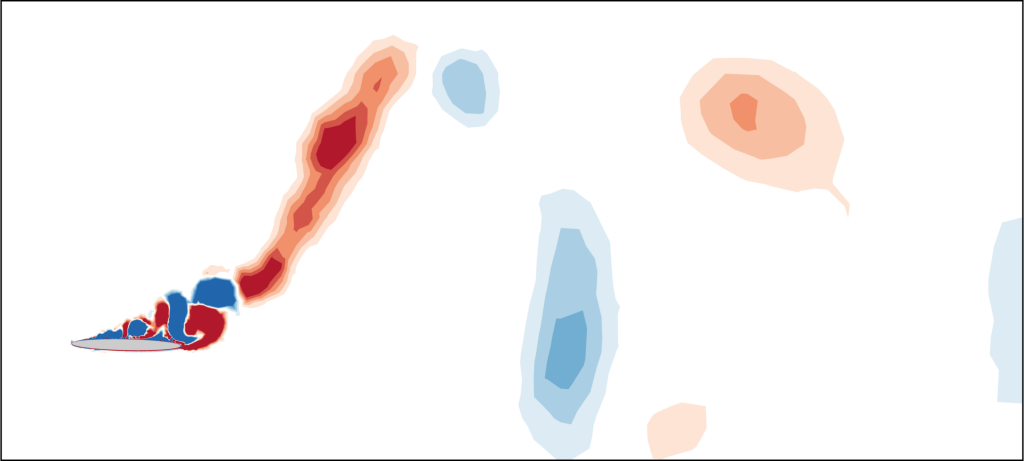}
        \caption{$ t/T \approx 0 $}
    \end{subfigure}\hspace{.5 cm}
    \begin{subfigure}{.4\textwidth}
        \includegraphics[width=\textwidth]{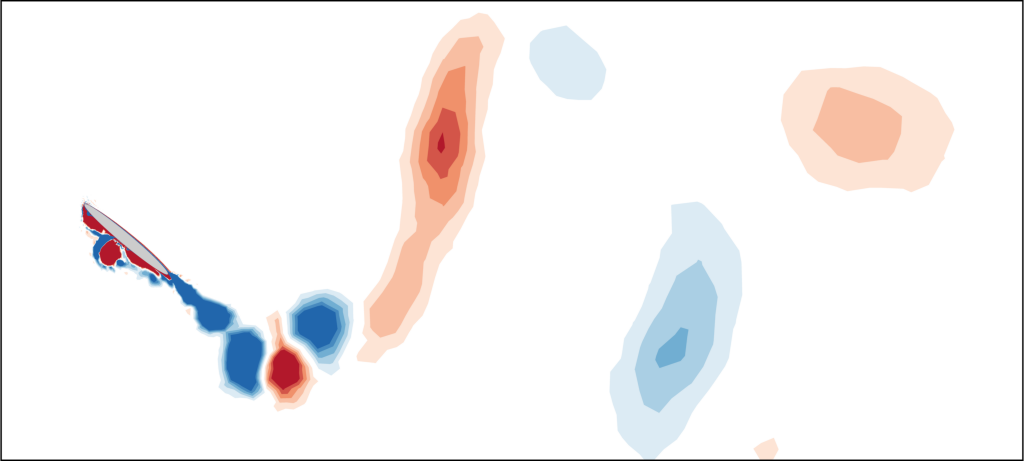}
        \caption{$ t/T \approx 0.25 $}
    \end{subfigure}
  \caption{Regime C vorticity contours demonstrated with kinematics $ h_0/c = 1.0, \theta_0 = 40^\circ, St = 0.533 $.}
  \label{fig:regimeC}
\end{figure}

\begin{figure}
  \centering
    \includegraphics[height=0.75cm]{colormap_3-eps-converted-to.pdf}

    \begin{subfigure}{.4\textwidth}
        \includegraphics[width=\textwidth]{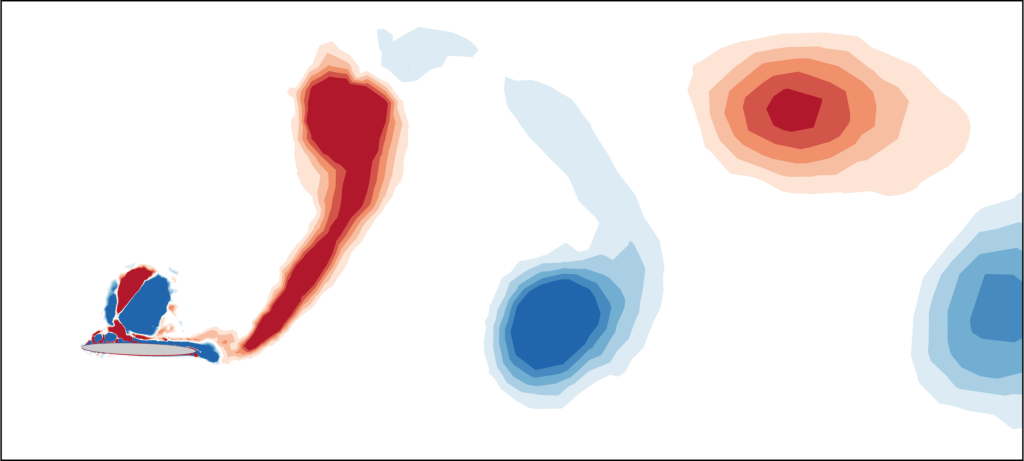}
        \caption{$ t/T \approx 0 $}
    \end{subfigure}\hspace{.5 cm}
    \begin{subfigure}{.4\textwidth}
        \includegraphics[width=\textwidth]{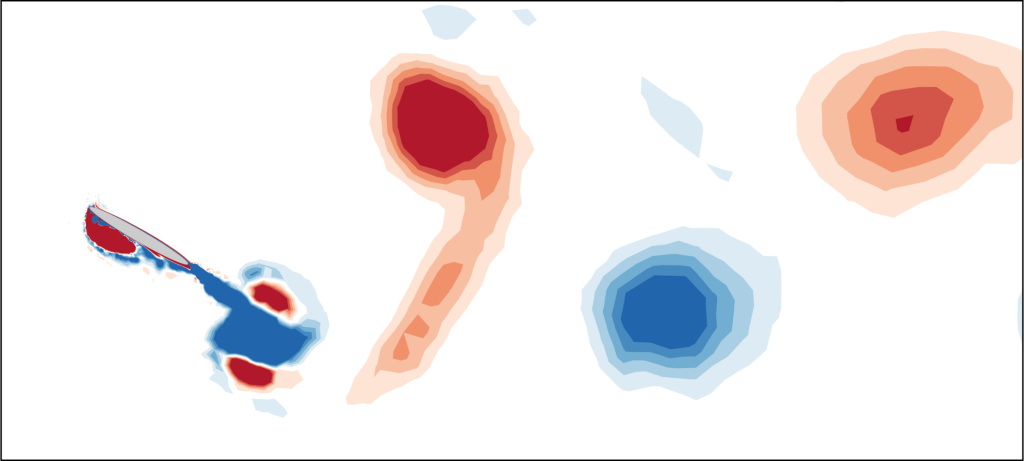}
        \caption{$ t/T \approx 0.25 $}
    \end{subfigure}
  \caption{Regime D vorticity contours demonstrated with kinematics $ h_0/c = 1.0, \theta_0 = 30^\circ, St = 0.6 $).}
  \label{fig:regimeD}
\end{figure}

Simulations with $h_0/c=1$ have been classified into four different flow regimes based on whether flow separation occurs, the number of vortices generated in the wake, and their position. The characterization of the wake regimes is also analyzed in relation with the performance indicators of thrust and efficiency. To observe the differences between these regimes, contour plots of normalized vorticity for four sets of kinematics are included in figures \ref{fig:regimeA}, \ref{fig:regimeB}, \ref{fig:regimeC} and \ref{fig:regimeD}, demonstrating the vortex structures on the foil and in the near wake at the bottom of the stroke, and at mid-upstroke. 



Regime A includes the kinematics for which there is minimal flow separation, which occurs at low frequency and very low angles of attack ($\alpha_{max} < 10^\circ$). These cases have the highest efficiency but lowest thrust output due to the low angle of attack.
As observed in figure \ref{fig:regimeA}, very weak vortices are shed at the extremes resulting in a ``2P" wake pattern but the trailing wake is mostly dominated by a long continuous streak of low magnitude vorticity alternating in sign for upstroke and downstroke.


In contrast to regime A, regime B does exhibit flow separation mid-chord, resulting in a slightly more coherent ``2P" shedding mode as observed in figure \ref{fig:regimeB} where a pair of vortices of opposite signs can be seen at the top and bottom of the stroke. However, one of the vortices in the pair gets dissipated much more quickly than the other, resulting in a reverse von K\'arm\'an type wake. This occurs for kinematics with moderate values of $\alpha_{max}$ between $10^\circ$ to approximately $17^\circ$. The thrust output is low ($C_T=0.25$ to $0.75$) and the efficiency is relatively high at 0.5 to 0.6.


\begin{figure}[t]
\centering
\begin{subfigure}{0.4\linewidth}
  \includegraphics[width=\textwidth]{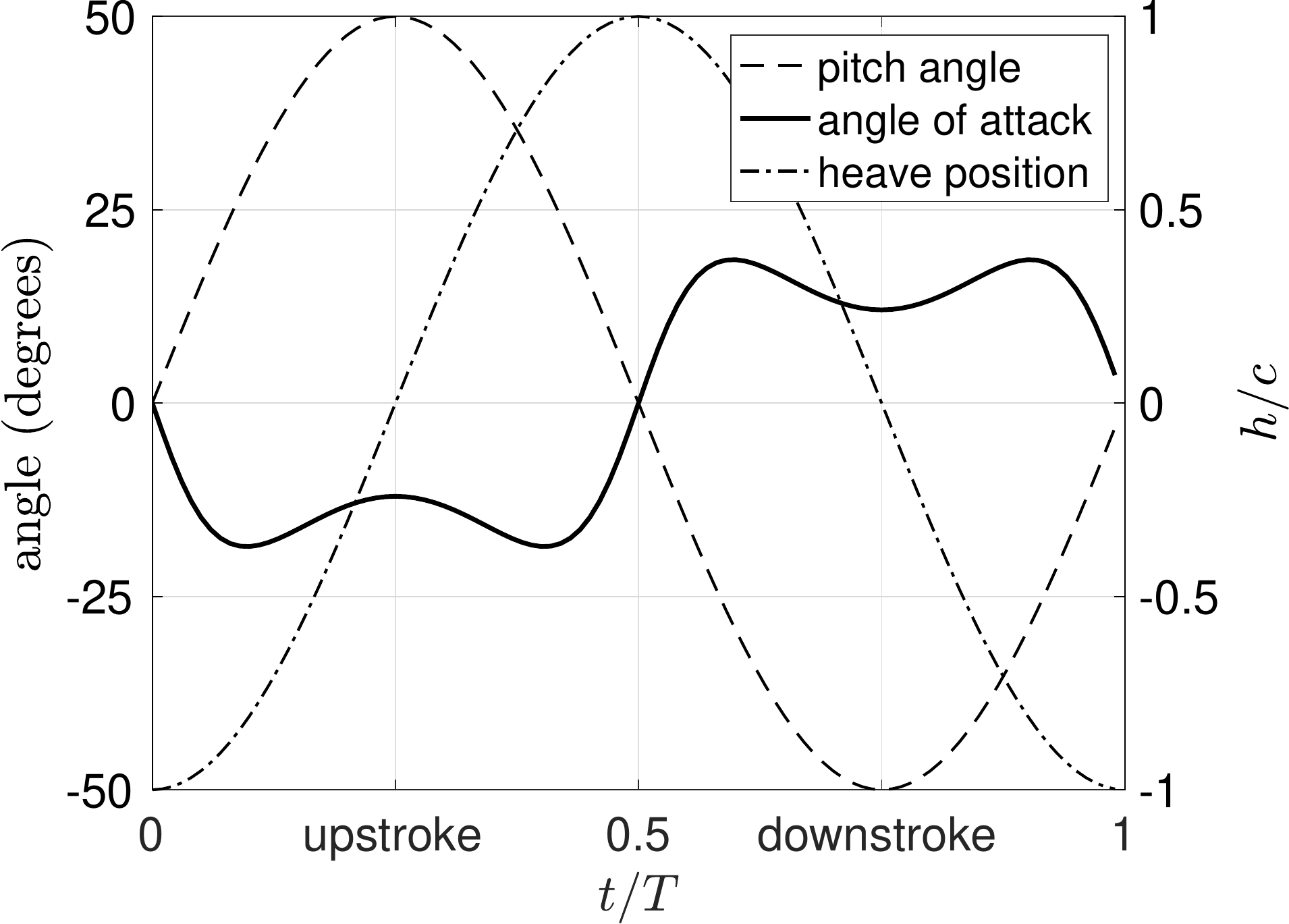}
  \caption{$ h_0/c = 1.0, \theta_0 = 50^\circ, St = 0.6 $ (regime C)}
  \label{subfig:kinematics_b}
\end{subfigure}\hspace{.5 cm}
\begin{subfigure}{0.4\linewidth}
  \includegraphics[width=\textwidth]{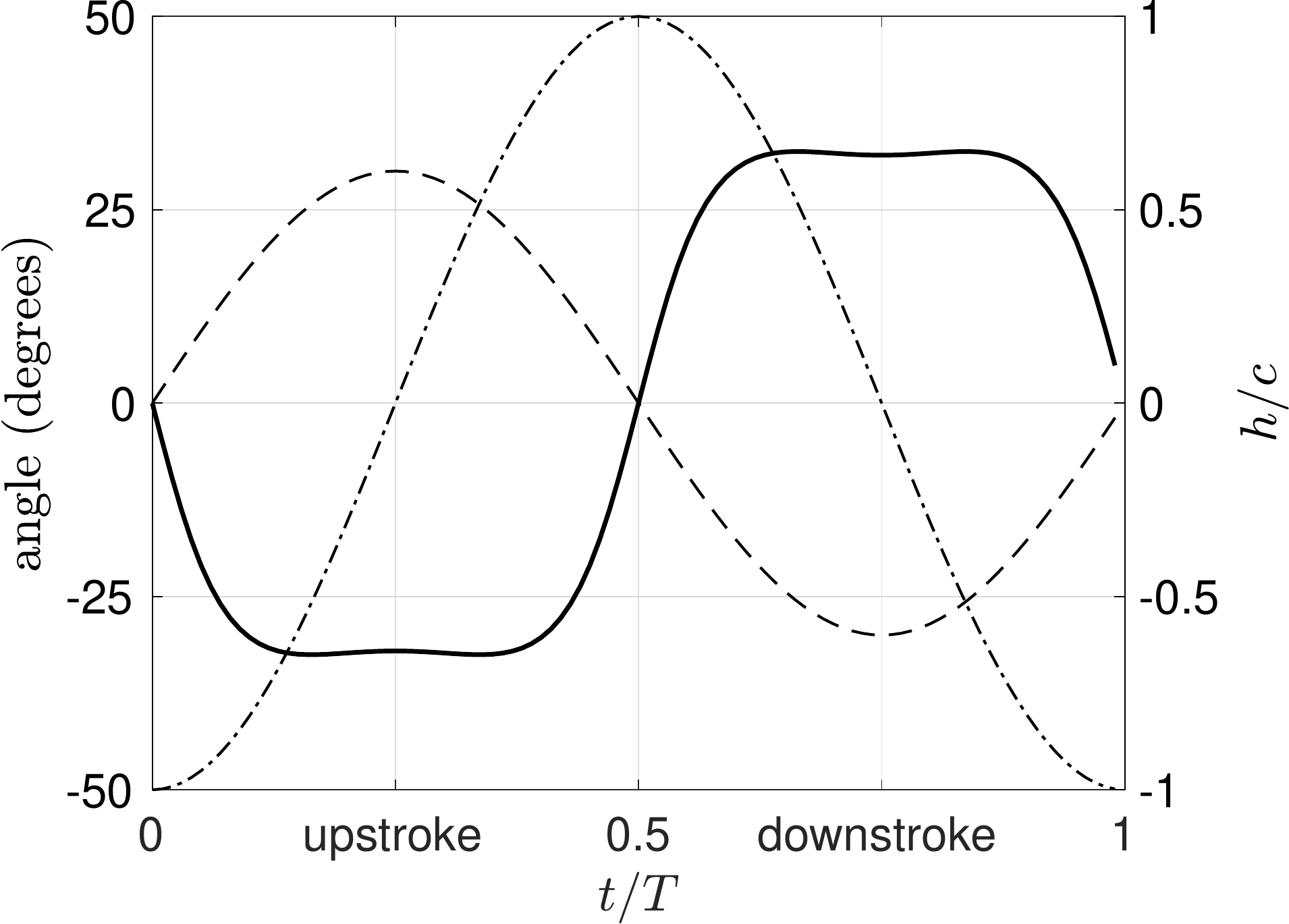}
  \caption{$ h_0/c = 1.0, \theta_0 = 30^\circ, St = 0.6 $ (regime D)}
  \label{subfig:kinematics_a}
\end{subfigure}

\caption{Change in heave position, pitch angle and angle of attack with oscillations for two different kinematics.}
\label{fig:kinematics}
\end{figure}

Regime C encompasses the kinematics for which strong flow separation occurs but a clear jet profile is not yet created in the wake. This regime is associated with moderately high $\alpha_{max}$ from $18^\circ$ to approximately $27^\circ$. Due to high pitch angles the maximum angle of attack does not occur at mid-stroke, but instead forms an ``M" profile as shown in figure \ref{subfig:kinematics_b}. This non-sinusoidal angle of attack variation during oscillations prevents the formation of a strong jet profile in the wake as there are multiple vortices shed each half-stroke with weaker vortices shed during mid-stroke, and slightly stronger ones shed at the stroke reversal. Three to four chord lengths downstream the stronger vortices prevail and a weak ``2S" pattern emerges, however closer to the foil the vortex pattern is more chaotic and a mix of ``2P" or ``P+S" wake patterns.


As seen in figure \ref{fig:regimeD}, regime D has only two primary vortices created per each cycle of oscillation. This is because the variation of angle of attack throughout the stroke has a flattened peak, as demonstrated by figure \ref{subfig:kinematics_a}. 
The kinematics for regime D have a high $\alpha_{max}$ greater than $28^\circ$. 
There is trailing edge vorticity created during the following stroke and other smaller structures that are much weaker than the strong primary vortices. This vorticity and small vortex structures are absorbed into the primary vortices, strengthening them further.
The vortex shed at top of the stroke is counter-clockwise whereas that on the bottom is clockwise in direction, creating a reverse von K\'arm\'an wake or a ``2S" vortex pattern. This results in an effective ``jet" motion and hence very high thrust generation ($C_T=0.75$ to $1.90$), but low propulsive efficiency ($\eta<0.4$).

For the higher heave amplitude of $h_0/c = 2$, all kinematics examined either result in a chaotic trailing wake (regime C) or semi-attached flow with very weak vortices in the wake (regime B). At high relative angles of attack of $\alpha_{max}>20^\circ$ there are inevitably multiple vortices generated per half-stroke as most of the cases have a distorted angle of attack profile similar to that shown in figure \ref{subfig:kinematics_b}. Moreover, vortices generated at the end of a stroke are entrained into middle of the wake because of the very high momentum of the foil and thus a clear reverse von K\'arm\'an wake (regime D) is rarely observed. At the other extreme, there are very few kinematics explored with $\alpha_{max}<10^\circ$. If more kinematics were tested it is possible that this region would result in attached flow, however it would likely produce minimal thrust. 

\subsection{Summary of thrust and efficiency performance at high heave amplitudes}

\begin{figure}[h!]
\centering
\begin{subfigure}{.45\textwidth}
  \centering
  \includegraphics[width=.9\linewidth]{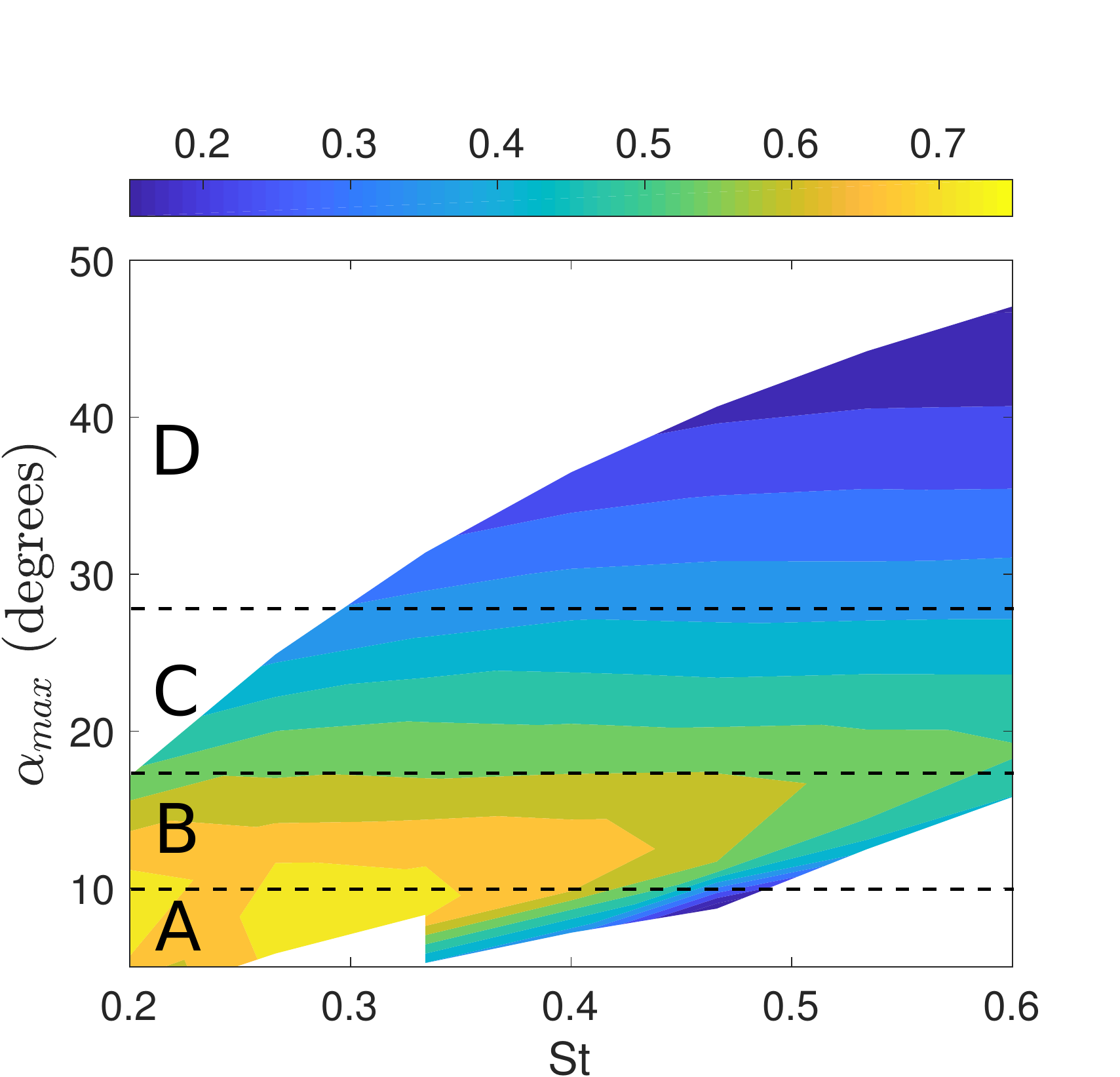}
  \caption{$\eta$ ($h_0/c = 1$)}
    \label{subfig:effh1}
\end{subfigure}%
\begin{subfigure}{.45\textwidth}
  \centering
  \includegraphics[width=.9\linewidth]{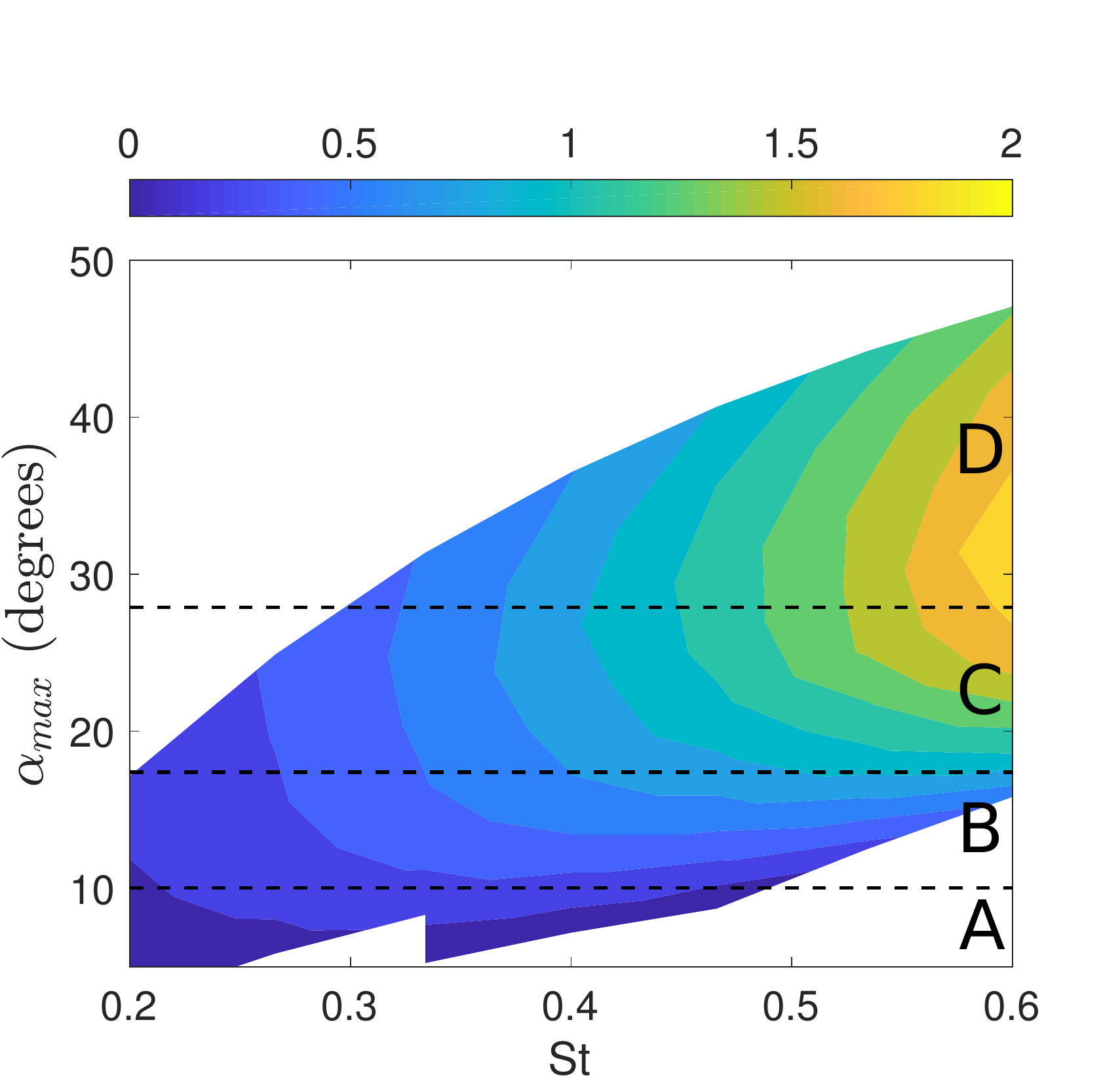}
  \caption{$C_T$ ($h_0/c = 1$)}
  \label{subfig:thrusth1}
\end{subfigure}

\par\bigskip
\begin{subfigure}{.45\textwidth}
  \centering
  \includegraphics[width=.9\linewidth]{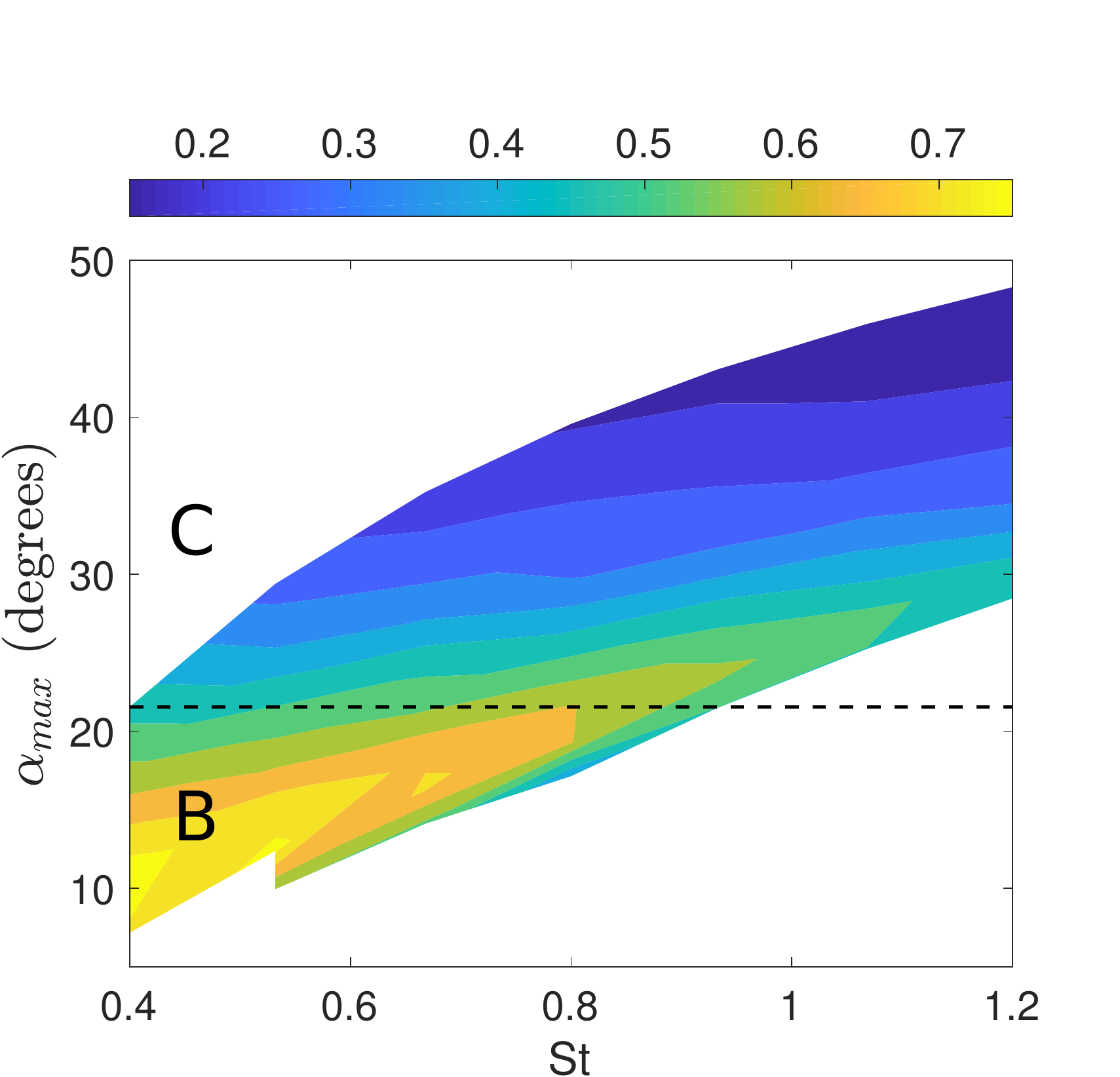}
  \caption{$\eta$ ($h_0/c = 2$)}
    \label{subfig:effh2}
\end{subfigure}%
\begin{subfigure}{.45\textwidth}
  \centering
  \includegraphics[width=.9\linewidth]{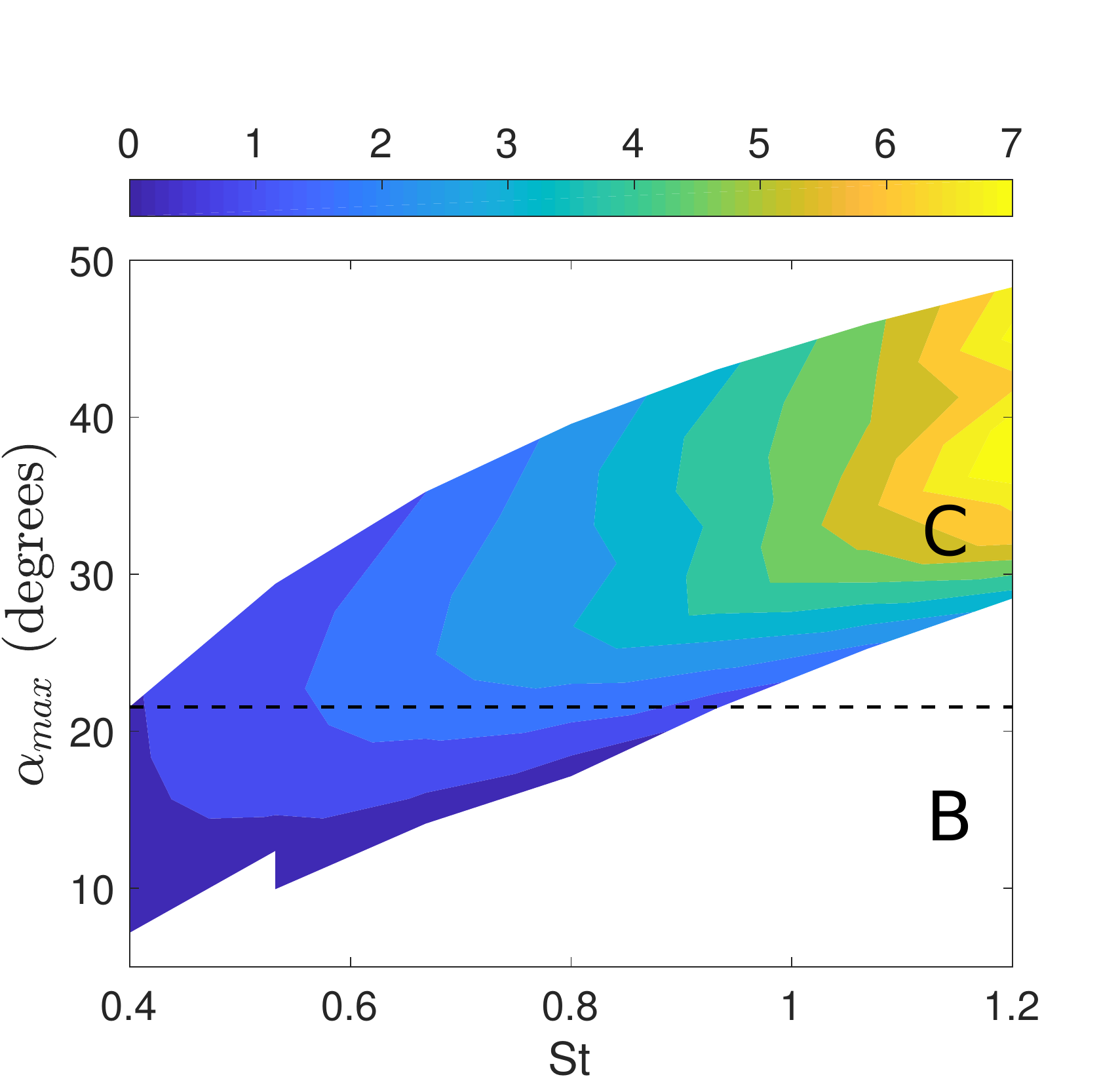}
  \caption{$C_T$ ($h_0/c = 2$)}
  \label{subfig:thrusth2}
\end{subfigure}
\caption{Contour plots of efficiency (left) and thrust coefficient (right) for the 2 sets of simulations, with the hydrodynamic regimes based on vortex patterns approximately delineated on the parameter space.}

\label{fig:all_plots}
\end{figure}

The thrust coefficient and efficiency are computed for each simulation and are presented in figure \ref{fig:all_plots} as a function of $St$ and $\alpha_{max}$, and for $h_0/c=1$ and $h_0/c=2$. The hydrodynamic regimes based on vortex patterns discussed in section \ref{subsec:hydrodynamics} are superimposed on the parameter space. The angle of attack profile is highly distorted at high Strouhal numbers for $h_0/c=2$, and the maximum effective angle of attack $\alpha_{max}$ is much higher than the effective angle of attack at mid-stroke, $\alpha_{mid}$. As a result of this, the contour plots in figures \ref{subfig:effh2} and \ref{subfig:thrusth2} cover a narrower range of $\alpha_{max}$ as compared to figures \ref{subfig:effh1} and \ref{subfig:thrusth1}. 

Similar to previous OFP experiments and simulations, there is a trade-off between high thrust and high efficiency as is also the case with other propulsion devices.
For both heave amplitudes, there is a maximum efficiency of approximately 75\% that occurs at the lower $St$ range of $St=0.3$ for $h_0/c=1$ and $St=0.4$ for $h_0/c=2$. The maximum efficiency occurs when the relative angle of attack is less than 10 degrees, corresponding to a fully attached boundary layer throughout the stroke (regime A) for $h_0/c=1$ and semi-attached (regime B) for $h_0/c=2$.  However at these kinematics, the thrust coefficient is still relatively low ($C_T < 0.5$).  As the $St$ increases and as $\alpha_{max}$ increases, the efficiency drops, but the thrust coefficient increases, reaching a maximum around $\alpha_{max}=\numrange{30}{40}$ degrees (regime C/D), and $St=0.6$ for $h_0/c=1$ and $St=1.2$ for $h_0/c=2$.

Although the Strouhal number with maximum thrust coefficient is twice as high for $h_0/c=2$ in figure \ref{subfig:thrusth2}, it is at the same non-dimensional frequency, $fc/U$, as the maximum thrust coefficient in figure \ref{subfig:thrusth1}. Since the higher heave amplitude intercepts a larger flow area, it is expected that higher thrust coefficients are reached in figure \ref{subfig:thrusth2}, with a maximum value of $C_T=8.22$, compared with the lower heave of $h_0/c=1$ where the maximum value is $C_T=1.97$. Using an alternative definition in equation \ref{eqn:CTstar}, the thrust coefficients are normalized with respect to their intercepted flow area. 
For $h_0/c=2$, that would result in the relation $C_T^* = C_T/4$, while for $h_0/c=1$, the relation would be $C_T^* = C_T/2$. Hence the thrust coefficient normalized by intercepted flow area is still higher at a higher heave amplitude for the same set of frequencies. This trend is expected to continue as the heave amplitude is increased. However it is hypothesized that the stresses on the foil and other mechanical constraints during operation may become a limiting factor.

The cumulative results at high heave ($h_0/c=\numrange{1}{2}$) give insight into how the efficiency and thrust coefficient values change with variation of the kinematic stroke, which is pertinent to OFP design considerations. 
For example, consider an oscillating foil of dimensions $c=0.3$ \si{m} and span of $1$ \si{m} to power a hypothetical small transport vessel. When operating at $h_0/c=1$, the total swept height is $2h_0=0.6$ \si{m}, corresponding to a swept area of $A=0.6$ \si{m^2}.

To assess the thrust performance at low speed (typical of a Bollard pull condition), a simulation at $St=2.0$ and $\theta_0=60^\circ$ yielded a thrust coefficient of $C_T=13.8$  ($C_T^*=6.9$) and $\eta=19\%$. This corresponds to an average thrust of $2.13$ \si{kN} at $U=1$ \si{m/s}, and an oscillation frequency of 3.3 \si{\hertz} for the above mentioned foil. As the vessel speed increases, the frequency and pitch amplitude can be constantly changed to improve efficiency, albeit at lower thrust coefficient. For example at $St=0.6$ and $\theta_0=50^\circ$, the thrust coefficient is $C_T=1.1$ ($C_T^*=0.54$) and the efficiency has improved to 54\%. At a speed of $U=5$ \si{m/s}, this corresponds to $4.2$ \si{kN} of thrust with an oscillation frequency of $5$ \si{\hertz}.

The scenario above is just one example of traversing the kinematics to achieve desired performance metrics by modifying pitch and frequency. Depending on the design requirements and the OFP control mechanism, variation in heave amplitude can also be dynamically implemented to satisfy transient yet high thrust requirements.

\section{Conclusions} \label{sec:conclusions}

	RANS simulations are performed at high Reynolds number ($Re=10^6$) to investigate the performance of oscillating foil propulsion (OFP). Motivated by the desire to explore OFP technology on surface vessels, a wide parameter space is explored in terms of frequency and pitch amplitudes for an elliptic fore-aft symmetric foil at high heave amplitudes of $h_0/c=1$ and $h_0/c=2$. The computed thrust coefficient and propulsive efficiency are reported and correlated with the vortex dynamics in the near wake.
	
	To assess the effects at high Reynolds number, the RANS results are compared with DNS results at a low Reynolds number ($Re=10^3$) at $h_0/c=1$. The propulsive efficiency at high Reynolds number are greater than those at low Reynolds number due to a higher lift-to-drag ratio and more resistance of the turbulent boundary layer towards separation. This difference emphasizes the importance of using an appropriate Reynolds number for OFP predictive models.
	
	The kinematics for $h_0/c=1$ are classified into four hydrodynamic regimes based upon flow separation and vortex patterns in the wake and analyzed with respect to the performance results. The kinematics with high angles of attack result in either a reverse von K\'arm\'an wake with exactly two distinct and strong vortices created per cycle, or a wake with chaotic vortex patterns, often due to a distorted non-sinusoidal angle of attack profile, representing sub-optimal thrust generation. For low angles of attack, the highest propulsive efficiency and lowest thrust is described by fully attached flow. Moderate angles of attack produce a semi-attached flow regime forming weak vortices.
	
	The flow dynamics for $h_0/c=2$ included only two of the four regimes observed for $h_0/c=1$, as a clear reverse von K\'arm\'an wake is not observed due to high heave and momentum of the foil and the non-harmonic angle of attack profiles. On the other extreme, a fully attached regime is not observed for the kinematic range explored at $h_0/c=2$.  
	
	The computational methods utilized in this paper are two-dimensional, and although the trends in terms of power and thrust align well with similar experimental studies, the three-dimensional effects due to vortex break-up or tip effects are not captured. However due to the lower computational cost, the current investigation is able to broadly sweep 126 unique high-heave kinematics. Future simulations, narrower in scope, can elucidate more details on the flow physics in these high-heave regimes by using three-dimensional models such as LES or RANS.
	
	Although propulsion from oscillating foils has been previously investigated computationally and experimentally, the presence of coupled high-heave and pitch kinematics is lacking in the literature, particularly at high Reynolds numbers. The kinematics explored here at $h_0/c=1$ and $h_0/c=2$ display very high thrust coefficients, up to $C_T=8.22$.  Although this high thrust comes at a decreased propulsive efficiency, vessels with OFP can transition to this high-thrust regime in short duration then transition back to high efficiency regimes with a simple modification to the kinematic stroke. The broad sweep of kinematics explored in this study can lay out the foundation for such design and control models.

\section*{Acknowledgements}
The authors acknowledge funding through a Rhode Island Commerce Corporation Innovation Voucher, and are grateful for the technical input from Tom Derecktor and Steve Winckler at Blusource Energy Inc. The research was conducted using computational resources and services at the Center for Computation and Visualization, Brown University.


\bibliography{propulsion}

\end{document}